\documentclass[10pt]{article} 
\usepackage{graphicx}
\usepackage{multicol}        
\usepackage{hyperref}
\usepackage{amssymb}
\usepackage{amsmath}
\usepackage{array}
\usepackage{graphics}
\usepackage{graphicx}
\usepackage{array}
\usepackage{fancyhdr}
\usepackage{enumerate}
\usepackage{floatrow}

\newcommand{\cm}{{\cal M}}

\newcommand{\parder}[2]{\frac{\partial {#1}}{\partial {#2}}}

\usepackage{authblk}
\title{Searches for magnetic monopoles and\\others stable massive particles}
\author[1,2]{Maurizio Spurio}
\affil[1]{\scriptsize{INFN - Sezione di Bologna, Viale Berti-Pichat 6/2, 40127 Bologna, Italy}}
\affil[2]{\scriptsize{Dipartimento di Fisica e Astronomia dell'Universit\`a, Viale Berti Pichat 6/2, 40127 Bologna, Italy}}
\begin{document}
\maketitle

\begin{abstract}
The Standard Model (SM) of the microcosm provides an excellent description of the phenomena of the microcosm, with the triumph of the discovery of the Higgs boson. There are many reasons, however, to believe that the SM is incomplete and represents a valid theory at relatively low energies only. Of particular interest are the models based on complete symmetries, such as those attempting a true unification between leptons and quarks in terms of a single symmetry group (Grand Unified Theories, GUTs) and those attempting unification between fermions and bosons, such as the supersymmetry. 
This chapter is devoted to the description of stable and massive particles not predicted within the SM, their energy loss mechanisms and their searches in the cosmic radiation. The stability of these particles means that if they were produced at any time in the thermal history of the Universe, they would still be present as relic particles. Examples of stable massive particles discussed in this chapter include magnetic monopoles, strange quark matter and supersymmetric particles. In particular, we focus on the status of searches for magnetic monopoles (also inducing proton-decay processes), nuclearites and Q-balls in neutrino telescopes.\footnote{Prepared for the forthcoming book "Particle Physics with Neutrino Telescopes", C. P\'erez de los Heros, editor, (World Scientific)}
\end{abstract}


\section{Introduction}
\label{sec:Introduction}

This chapter is devoted to the description of stable massive particles not predicted within the Standard Model, and their searches in the cosmic radiation, in particular in neutrino telescopes. 
Stable massive particles are sufficiently long lived that they can be directly observed via strong and/or electromagnetic interactions in a detector rather than via their decay products. 
Their stability means that if they were produced at any time in the thermal history of the Universe, they would still be present as \textit{relic particles}. 
The motivation for the existence of relic massive objects is usually connected with general considerations on cosmology and dark matter.

The searches of different kinds of stable massive particles are an important component in the programs of collider experiments \cite{fair07}.
The searches of relic particles is also a fundamental aspect of many astroparticle physics experiments in space, on the Earth surface and underground/ice/water \cite{ms}.
Non-accelerator experiments are sensitive to masses many orders of magnitude larger than the TeV scale reachable (for instance) at the LHC collider. 

Examples of stable massive particles include magnetic monopoles, strange quark matter and supersymmetric particles. 

Magnetic monopoles ($\cm$s, in the following) are particularly intriguing particles, related not only to the problem of the non-observed dark matter, but also to the inner symmetries of electromagnetic interactions, \S \ref{sec:classical}.
The interest on $\cm$s raised with advent in the 1980's of Grand Unification Theories (GUTs), \S \ref{sec:GUT}, which have the objective to merge into a single unified interaction the strong and electroweak interactions at very high energies.  
The existence of $\cm$s and the proton decay are two predictions of GUTs that can be experimentally tested. 
GUT's monopoles (as other stable massive particles) can only be produced in an early epoch of the Universe history. The processes of $\cm$ production, adiabatic energy loss and their successive re-acceleration after the creation of large scale galactic magnetic fields are described in \S \ref{sec:hysto}.
A relevant aspect of the physics of $\cm$s is their energy loss through interaction with matter, described in \S \ref{sec:loss}. These mechanisms lead to different detection techniques used by different experiments to set upper limits on their flux, \S \ref{sec:searches}.

Strange quark matter (SQM), composed of comparable amounts of $u,\ d$ and $s$ quarks, is a hypothesized ground state of hadronic matter. Nuggets of SQM formed after the Big Bang could be survived up to now and be present in the cosmic radiation, \S \ref{sec:sqm}. Those with masses $\gtrsim 10^7$ the proton mass are called \textit{nuclearites}, and can be detected by large area experiments, as neutrino telescopes. 
Q-balls are another example of relic, stable massive particles made of aggregates of squarks $\tilde{q}$, sleptons $\tilde{l}$ and Higgs fields, \S \ref{sec:qballs}. Their characteristic is to produce in a detector a signature similar to that of a $\cm$ inducing proton decay processes. 

Finally, in \S \ref{sec:searchesNT}, we focus on the status of searches for $\cm$s in neutrino telescopes  (also inducing proton-decay processes), nuclearites and Q-balls. This is an active field, which could be subject to significant future improvements also with the present generation of experiments.   

\section{Dirac (or classical) Magnetic Monopoles}
\label{sec:classical}

James Clerk Maxwell developed his theory of electromagnetism \cite{maxwell} based on experimental observation of electric and magnetic interactions and on the Faraday's concept of \textit{field}. 
The Maxwell's theory of the electromagnetic fields was a fundamental breakthrough: it not only unified magnetism and electricity in one simple theory, but also explained the properties of light, which it was showed to be an electromagnetic wave. 
Furthermore, the theory derived numerically the value of the speed of light, demonstrating that is constant, and therefore paved the way for theory of relativity.
Although Maxwell's equations do not include magnetic charges, its structure allows embedding them easily, as shown in \S \ref{sec:Maxwell}

The compatibility of magnetic monopoles with quantum mechanics was due in 1931 to P.~A.~M. Dirac \cite{dirac}, who introduced the today called \textit{classical magnetic monopole}. Quantum theory allows the Dirac monopole under the condition of a particular minimum value of the magnetic charge, $g_D$. The interesting aspect of the Dirac theory is that the existence of a free magnetic charge $g_D$ would explain the quantization of the electric charge, $e$. 
Dirac established the basic relation between $e$ and $g$ as:
\begin{equation}\label{eq:1.g}
{eg\over c}={n\hbar \over 2 } \longrightarrow g = n\cdot g_D = n\cdot {1\over 2} {\hbar c \over e} \sim n\cdot {137\over 2} e \ .
\end{equation}
where $n$ is an integer. This condition is derived in \S \ref{sec:Dirac}

\subsection{The symmetric Maxwell equations}\label{sec:Maxwell}

The equations of Maxwell become symmetrical in form with the introduction of magnetic charges $g$, with density denoted as $\rho_m$ and density current as $\vec{j}_m$. In the Gauss CGS symmetric system of units, one has:
\begin{align}  \label{eq:mx}
  \begin{aligned}
    \vec{\nabla} \times \vec{E} &=\frac{4\pi}{c}\,\vec{j}_m -\frac{1}{c}\parder{\vec{B}}{t},&
    \vec{\nabla}\cdot \vec{B} &= {4\pi\rho_m} ,
    \\
    \vec{\nabla}\times \vec{B} &=\frac{4\pi}{c}\,\vec{j}_e + \frac{1}{c}\parder{\vec{E}}{t}
    ,&
    \vec{\nabla}\cdot \vec{E} &=4\pi\rho_e.
  \end{aligned}
\end{align}
The familiar equations studied at the University courses are recovered when $\rho_m=0$ and $\vec{j}_m=0$. 
In vacuum (i.e. when the electric density $\rho_e=0$ and density current $\vec{j}_e=0$) the equations have a symmetry known as \textit{electric-magnetic duality}: by replacing the electric and magnetic fields as
\begin{equation}\label{eq:dua}
\vec{E}\rightarrow \vec{B}, \textrm{    and   } \vec{B}\rightarrow - \vec{E} 
\end{equation}
the equations remain unchanged. 
The presence of the electric charge terms, $\rho_e, \vec{j}_e$, destroy the electric-magnetic duality, which is restored if we assume the existence of magnetic charges. 

From Eq. (\ref{eq:mx}), electric and magnetic fields applied to an electric $e$ or magnetic charge $g$ give rise to the Lorentz forces:
\begin{equation}\label{eq:fe}
\vec{F}_e = e \biggl(\vec{E} + {1 \over c} \vec{v}_e \times \vec{B}\biggr) \ ,
\end{equation}
\begin{equation}\label{eq:fg}
\vec{F}_g = g \biggl(\vec{B} - {1 \over c} \vec{v}_m \times \vec{E}\biggr) \ .
\end{equation}
This beautiful symmetry is experimentally broken since the electric charges exist but magnetic charges have never been observed. 
From the point of view of classical electrodynamics, there is no fundamental reason to exclude the existence of magnetic charges, i.e. Maxwell's equations are perfectly compatible with magnetic monopoles and their existence would make the theory more symmetric.

If a magnetic charge $g_D$ exists, the Dirac quantization relation \ref{eq:1.g} automatically explains the quantization of the electric charge, i.e. the electric charge of any particle is an integer multiple of the elementary charge, $e$.
This is a remarkable prediction, because electric charges are indeed quantized and integer multiples of the electron charge $e$ (or one third of it, $e/3$, if one includes quarks).
In any case, also for a different value of the minimum electric charge, eq. \ref{eq:1.g} holds for a different minimum $g_D$. 

Contrarily to classical electromagnetism, magnetic monopoles seem to be incompatible with quantum mechanics. 
In classical physics, the description in terms of potentials is in some sense optional, because observable quantities do not depend on the potential. 
By contrast, in quantum physics the potentials couple directly to the quantum wave function, with real physical consequences, as in the famous case of the Aharonov-Bohm effect \cite{ab}. The use of the $\phi, \vec{A}$ cannot be avoided, as discussed in the next section.

\subsection{The Dirac quantization of the $e,g$ charges}\label{sec:Dirac}

In quantum mechanics, electromagnetic interactions are described in terms of scalar and vector potentials, $\phi$ and $\vec{A}$, respectively.
The electric and magnetic fields are determined through the relations:
\begin{eqnarray}
\vec{E} &=& -\frac{1}{c}\frac{\partial \vec{A}}{\partial t}-\vec{\nabla}\phi,\nonumber\\
\vec{B} &=& \vec{\nabla}\times\vec{A}.
\label{eq:potentials}
\end{eqnarray}
In this representation, the duality symmetry between electric and magnetic forces seems to be broken. However, the potentials themselves are not physical, observable quantities: any transformation of this kind:
\begin{equation}\label{eq:gauge}
\vec{A} \Rightarrow \vec{A}^\prime = \vec{A} + \vec\nabla \chi \quad ,
\quad
\phi \Rightarrow \phi^\prime =\phi - \frac{\partial \chi}{\partial t}
\end{equation}
gives rise to the same electric and magnetic fields.
Changing from one of these physically equivalent potential configurations to another under Eq. \ref{eq:gauge} is known as a {\it gauge transformation}. Because physical quantities do not change under such gauge transformations, we say that the theory has a {\it gauge symmetry}. The gauge symmetry of Maxwell's equations is known in group theory as U(1).

The Standard Model of particle interactions is based on similar and more complex gauge symmetries. Gauge symmetries are thus the fundamental principle that determines the nature of particle interactions.

The definition of physical fields through Eq. \ref{eq:potentials} apparently forbid the existence of point-like magnetic charges. This is due to simple properties of vector calculus 
$\vec{\nabla} \cdot \vec{B}=\vec{\nabla}\cdot  (\vec{\nabla}\times\vec{A})= 0$.
In terms of field lines, magnetic field lines can never have end points and must be closed. 

In quantum mechanics, a particle is described in terms of a complex wave function $\psi$, whose square value $|\psi|^2$ in a given volume yields the probability of finding the particle in this particular volume. 
The complex nature of the wave function becomes apparent in interference experiments.
In the formalism of quantum electrodynamics (QED), $\vec{A}$ is seen as the space component of a four-vector, $A_{\mu }$, and $\phi$ the time component of the same four-vector. Electromagnetic interactions are adequately accounted for by the \textit{minimal coupling} formalism. 
Mathematically, minimal coupling is achieved acting on the four-momentum $ p_\mu$ of the free Lagrangian (or in the free-field equation, as for instance in the Dirac equation) of an electrically charged particle $e$ by the replacement: 
\begin{equation}
p_{\mu } \Rightarrow p_{\mu }-e\ A_{\mu } \ .
\end{equation}
The use of the potential instead of the magnetic field seems to rule out the hypothesis of the existence of magnetic charges, as the vector potential cannot describe a magnetic monopole term.
Nevertheless, Dirac in its paper of 1931 \cite{dirac}, showed that magnetic monopole solutions are, under certain conditions, still allowed. We try to reproduce the Dirac reasoning using a modern approach. 

\begin{figure}[tb]
\begin{center}
\includegraphics[width=8.0cm]{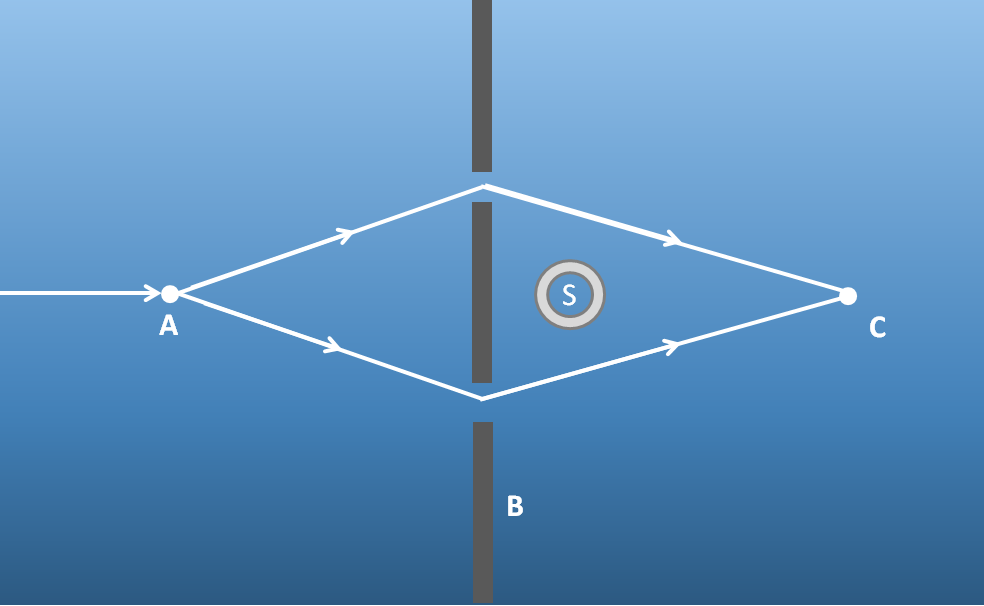}
\end{center}
\caption{\small\label{fig:2splits} Illustration of the Aharonov-Bohm experiment. See text for details}\label{fig:twosplits}
\end{figure}

Let us first consider the Aharonov-Bohm effect \cite{ab} as illustrated in Fig. \ref{fig:twosplits}. 
Charged particles emitted by the source A pass through the two slits in the screen B and are detected at C. Without the presence of the solenoid S, the amplitudes for the passage through the individual slits combine coherently and the probability density at C is given by
$P=|\psi_1+\psi_2|^2$, where $\psi_1$ is the probability for passage through the first slit and $\psi_2$ is that for the passage through the second one. If the solenoid S is placed between the two slits, the probability density at C becomes 
\begin{equation}\label{eq:ab}
P^\prime=|\psi_1+ \exp\biggl[\frac{ie\Phi_B}{\hbar c} \biggr]\cdot \psi_2|^2 \ ,
\end{equation}
where $e$ is the electric charge of the particles emitted in A and $\Phi_B$ is the magnetic flux through the solenoid. 
By moving the solenoid and observing the change in the interference pattern, one could detect the presence of the solenoid, unless
\begin{equation}\label{eq:abd}
\exp \biggl[\frac{ie\Phi_B}{\hbar c} \biggr] = 1 \ .
\end{equation}
This is the \textit{handhold} condition in the Dirac theory for the introduction of magnetic charges.

\begin{figure}[tb]
\begin{center}
\includegraphics[width=10.0cm]{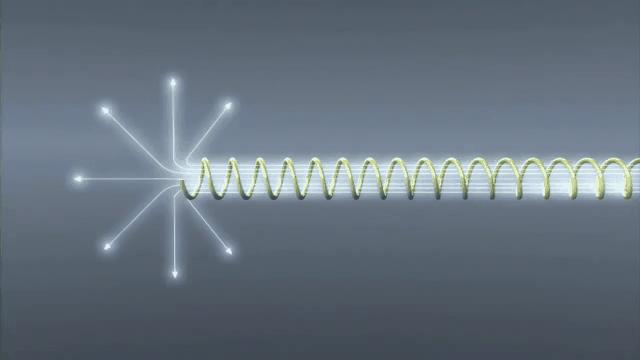}
\end{center}
\caption{\small\label{fig:diracM} Illustration of the field produced by a point magnetic pole and of the string, which one finds with the use of the vector potential. An animated version is available on  \url{https://www.youtube.com/watch?v=Et5fBRMHyMo}.  Credit: The MoEDAL experiment }\label{fig:diracm}
\end{figure}

In the Dirac's model, each magnetic South pole is connected to a magnetic North pole by a line of singularity called a \textit{Dirac string}, as shown in Fig. \ref{fig:diracm}.
That string corresponds to an idealized solenoid with thickness tending to zero, carrying magnetic flux from the South pole to the North pole. 
In this way, the magnetic field lines remain continuous (as for an electric dipole, the field lines are then exiting from the North pole and entering in the South). The shape of the magnetic field around the end of the solenoid looks exactly like that of a $\cm$ with magnetic charge $g$.
In fact, according to the Gauss theorem in Eq. \ref{eq:mx}, the magnetic flux is equivalent to the presence of a magnetic charge $g$ equal to 
\begin{equation}\label{eq:gphib}
\Phi_B=4\pi g \ . 
\end{equation} 
Along the length of the solenoid, the magnetic field is confined inside it, apart from the monopole fields at the two ends. 
Assuming the solenoid infinitesimally thin and long,
an analytic formula for the vector potential can be written. However, this vector potential expression is singular along the Dirac string, an unpleasant situation for a physical object.
The situation is solved assuming that the solenoid S in Fig. \ref{fig:twosplits} corresponds to a Dirac string connecting the two poles.
As the vector potential $\vec{A}$ affects the complex phase of the particle's wave function, the Dirac string can give rise to interference effects and be \textit{detectable}. However, under the condition of Eq. (\ref{eq:abd}) and assuming the relation (\ref{eq:gphib}), we have:
\begin{equation}\label{eq:abdg}
\exp \biggl[\frac{ie (4\pi g)}{\hbar c} \biggr] = 1 \quad \Rightarrow \quad
\frac{4\pi eg}{\hbar c}= {2\pi n} 
\end{equation}
where $n$ is an integer number. This relation yields Eq. (\ref{eq:1.g}), which is known as the \textit{Dirac quantization condition}. 
Under this condition, no experiment can observe the presence of a Dirac string. Thus, the string is not real; it is a mathematical artefact.
Only the two poles at the ends of the string are observable, and physically they appear as two separate particles, two free $\cm$s.

Also if magnetic monopoles can be consistently described in quantum theory, they do not appear automatically (unlike the famous Dirac prediction of the existence of the positron). 
In addition, in the Dirac model the $\vec{B},\vec{E}$ duality is not perfect, since the unit magnetic charge is much larger than the electric unit charge. According to Eq. (\ref{eq:1.g}), numerically one has
\begin{equation}\label{eq:gd}
g_D = 68.5 e = 3.3\ 10^{-8} \textrm{ esu} \ .
\end{equation}

\section{Magnetic monopoles in extensions of the Standard Model}
\label{sec:GUT}

The two decades around 1970-1980 were the golden age for the formulation of the Standard Model of particle physics. 
The guiding principle was the concept of gauge symmetry, and although the gauge transformations involved in weak and strong interactions are more complicated than in quantum electrodynamics (QED), the structure of these theories is modelled according to QED. 

The first, successful, theoretical and experimental step of the Standard Model was to unify electromagnetism and the weak interactions into the {\it electroweak} interaction.
The theory, developed by S. Glashow, S. Weinberg, A. Salam and others, has two different gauge symmetry groups known as U(1) and SU(2). 
The U(1) group has the same structure as the gauge symmetry of Maxwell's equations; the SU(2) symmetry for the weak interactions is somewhat more complicated. In the following, we refer to the Glashow, Weinberg and Salam model with its group representation, SU(2)$\times$U(1). This theory does not require the presence of $\cm$s.

In a slightly different variant of the electroweak model, first proposed by Georgi and Glashow~\cite{Georgi:1972cj} and known as SO(3),  't~Hooft~\cite{'tHooft:1974qc} and Polyakov~\cite{Polyakov:1974ek} found that the presence of both electric and magnetic charges are necessarily required. 
The model differs, with respect to the SU(2)$\times$U(1), in the so-called Higgs sector.
The mass of the $\cm$s foreseen by the SO(3) symmetry group would be of the order of 100 GeV/$c^2$, determined by the energy scale associated with weak interactions. Experimental results disfavoured the Georgi-Glashow model for the electroweak theory with respect to the SU(2)$\times$U(1), as well magnetic monopole solutions with masses detectable at (relatively) low energy accelerators. 

However, the Standard Model has two separate sectors: one for the electroweak interactions (mediated by the $W^\pm$ and $Z^0$ vector bosons and the photon), and the quantum chromo-dynamics (QCD) interaction among coloured quarks. These interactions, mediated by gluons, are also called \textit{strong interactions} and denoted in group theory as SU(3). 
Some \textit{conservation rules} separate the two sectors of the Standard Model: the lepton and baryon numbers conservation, for instance. This means that decay processes as
\begin{equation}\label{eq:pdecay}
p \rightarrow \pi^0 e^+
\end{equation}
are forbidden. 

The possibility of an \textit{unification} of the two sectors of the Standard Model is the next, ambitious, step of theory and experiments. Georgi and Glashow itself realized that the SO(3) model could be used as a prototype of a theory unifying the SU(3) and the SU(2)$\times$U(1) sectors of the Standard Model in one Grand Unified Theory (GUT) \cite{Georgi74}. This GUT would describe all known elementary particle interactions, except gravity. 
The unification of the electromagnetic, weak and strong interactions would occur at very high energies, above $\sim 10^{15}$ GeV (the GUT energy scale).
One of the characteristic of GUTs is that the baryon and lepton numbers are not separately conserved: processes similar to that described in (\ref{eq:pdecay}) are allowed.
Below GUT energy threshold, the symmetry would be broken into the SU(2) $\times$ U(1) of the electroweak interactions and the SU(3) of strong interactions.

Different symmetry groups have been proposed as GUTs. The first and simplest one, the SU(5) proposed by Georgi and Glashow, predicted a proton decay lifetime (\ref{eq:pdecay}) in the range of $10^{30}-10^{31}$ years, well within experimental observations using kiloton detectors (in one kiloton there are more than $10^{32}$ nucleons). The prediction of a measurable proton decay lifetime was one of the first motivation for the construction of large underground experiments in the 1980s. So far, no proton decays have ever been observed \cite{pd}. 
Experimental results rule out the original Georgi-Glashow SU(5) GUT, but other GUTs predicting a much longer proton lifetime survive.   

Because the structure of the SU(5) was similar to the Georgi-Glashow electroweak model, it predicted 't Hooft-Polyakov $\cm$s as well, but with much higher masses, around $10^{16}$ GeV/c$^2$. 
However, the existence of $\cm$s is not specific to the SU(5). It was shown that is an unavoidable consequence of GUTs: any theory where a simple gauge group is spontaneously broken into a subgroup that contains an explicit U(1) factor would have 't Hooft-Polyakov monopoles \cite{pres84}. 

\begin{figure*}[tb]
\begin{center}
\includegraphics[width=11.0cm]{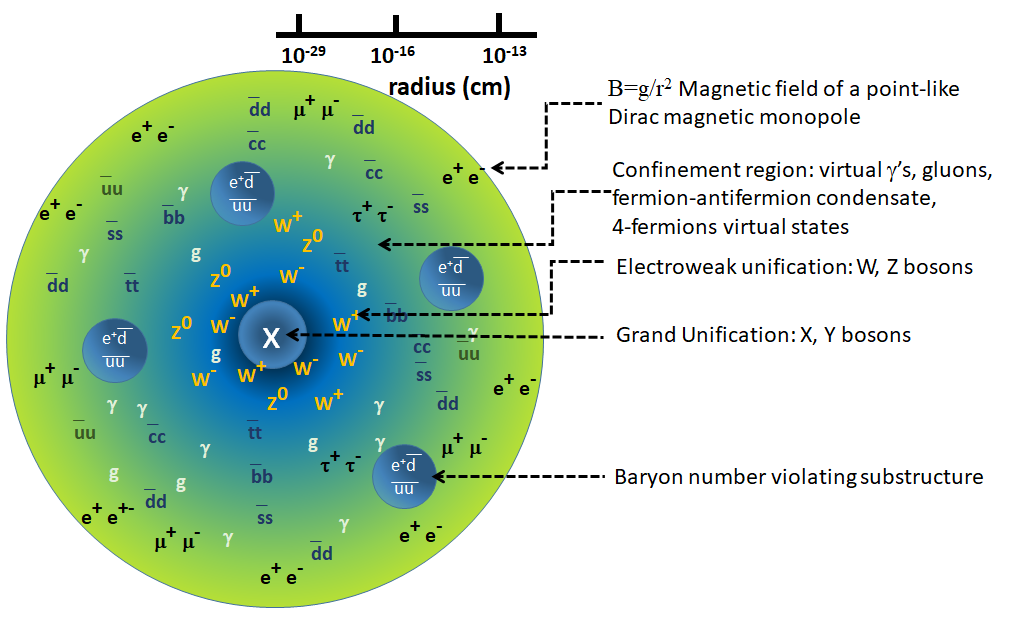}
\end{center}
\caption{\small\label{fig:gutM} Illustration of the GUT $\cm$ structure. The sketch shows various regions corresponding to 
$i)$ grand unification ($r\sim 10^{-29}$ cm): inside this core virtual $X$, $Y$ bosons should be present. They are the bosons allowing processes as the proton decay (\ref{eq:pdecay}); 
$ii)$ electroweak unification ($r\sim 10^{-16}$ cm) with virtual $W^\pm,\  Z^0$ bosons; 
$iii)$ the confinement region ($r\sim 10^{-13}$ cm) with virtual photons, gluons and a condensate of fermion-antifermion, four-fermion bags. 
For radii larger than few fm one has the field of a point magnetic charge, $B = g/r^2$.}
\end{figure*}

There is no prediction for the mass of the Dirac magnetic monopole, which is assumed as a structure-less, point-like particle. On the contrary, GUT $\cm$s are expected to be very massive, composite objects, as shown in Fig. \ref{fig:gutM}. The masses depends on the particular model, but are always above the GUT energy scale.
The central core of a GUT $\cm$ retains the original grand unification among interactions and contains the fields of the superheavy gauge bosons, which mediate baryon number violation. 
The fermion-antifermion condensate would have baryon number violating substructures that would allow the $\cm$ catalysis of the proton decay, see \S \ref{sec:cata}.

At GUT energies, it is obviously impossible to perform tests with accelerator experiments. The foreseen $\cm$, if they exist, would be a stable particle; it could be destroyed only by annihilation with another $\cm$ of opposite magnetic charge.
The search for relic $\cm$s is another possibility for experimentalists to test GUTs. Recent reviews on the different monopole models are in \cite{Milton,raja}.

Magnetic monopoles with Intermediate Mass $\sim 10^{5} \div 10^{13}$ GeV/c$^2$ (IM$\cm$) are predicted by theories with an intermediate energy scale between the GUT and the electroweak scales \cite{laza-imm}. 
The structure of an IM$\cm$ would be similar to that of a GUT monopole, Figure \ref{fig:gutM}, but without any term violating baryon number conservation. Monopoles with intermediate mass might have been produced in the early Universe \cite{keph01} and survived as relics; they would be stable and do not catalyse proton decay.
As IM$\cm$s would be produced after inflation, their number would not be reduced by inflationary mechanisms. Of particular relevance for experiments, IM$\cm$s with $10^{7} < M < 10^{12}$ GeV/c$^2$ could acquire nearly-relativistic velocities in one coherent domain of the galactic field \cite{bhatta}.
Experimentally one has to search for IM$\cm$s with $\beta \gtrsim 0.1$.

\section{The thermal history of the Universe and Monopoles}
\label{sec:hysto}

According to the Big Bang model, the Universe started in a state of extremely large density and very high kinetic energies (in the following of the section, we use the temperature as a scale of the kinetic energy through the Boltzmann constant $k$) of the initial plasma of particles. 
As time progressed, density and temperature decreased and the particle composition changed.
The GUT phase, with the unification of strong and electroweak interactions, lasted until the plasma temperature dropped to $ T_{GUT}\sim 10^{14}\div 10^{15}$ GeV. At this energy, a phase transition is thought to have occurred, and GUT $\cm$s with masses 
$M c^2 \gtrsim T_{GUT}/\alpha \simeq 10^{16}\div 10^{17}$ GeV appeared.
Here, $\alpha\simeq 0.025$ is the dimensionless unified coupling constant.

According to the production mechanisms of such topological defects in the early Universe, we can derive how many $\cm$s there should be in the Universe today. 
However, the predicted density number is far too high to be compatible with observations \cite{pre79}. 
The precise density today depends mainly on the symmetry breaking scale and some theoretical assumptions. Generic models \cite{kibble} for GUT $\cm$s suggest a relic monopole abundance density $\rho_M$ of \cite{burdin}
\begin{equation}\label{eq:aboun}
\Omega_m \equiv \frac{\rho_M}{\rho_c} \sim 10^{11} \biggl(\frac{T_{GUT}}{10^{14}\textrm{ GeV}} \biggr)^3 \biggl(\frac{M}{10^{16}\textrm{ GeV/c}^2}\biggr) \ ,
\end{equation}
where:
\begin{equation}\label{eq:2.rhoc}
\rho _{c}\equiv\frac{3H^{2}}{8\pi G_{N}}
= 1.88\cdot 10^{-29} { h}^{2}\ \textrm{ [g cm}^{-3}]
= 1.05\cdot 10^{-5} { h}^{2}\ \textrm{ [GeV cm}^{-3}] \ ,
\end{equation}
is the critical energy density of the Universe \cite{pdg}. In Eq. \ref{eq:2.rhoc} the scaled Hubble parameter, $h\sim 0.7$, is defined in terms of the Hubble constant $H\equiv 100 h$ km s$^{-1}$ Mpc$^{-1}$. 
Eq. (\ref{eq:aboun}) corresponds to a number density of $\sim 10^{-21}$ cm$^{-3}$.
This very large number is in contradiction with observational cosmology \cite{pdg}. 
This overabundance was known as the \textit{magnetic monopole abundance problem}, that needs a dilution factor of at least $10^{13}$ to be solved. The reduction of $\cm$s in the Universe was one of the motivating factors for cosmological inflation in Guth's original work \cite{guth}.

An inflationary phase in the evolution history of the Universe is completely in agreement with observational cosmology. A large number of theoretical models exist that give rise to enough inflation to solve the monopole abundance, horizon and flatness problems and can produce perturbations that are compatible with observations. However, the increase of the volume is too high, and GUT $\cm$ number density is reduced to a very small value. 
If inflation happens after monopoles are formed, in most models they are effectively diluted such that their abundance would be something of the order of one $\cm$ in our Universe.
Larger abundances can be obtained in scenarios with IM$\cm$s, or with carefully tuned parameters or if the reheating temperature was large enough to allow $\cm$s production in high-energy collisions of fermions $f$, like $f \overline f \rightarrow \cm \overline \cm$.

\subsection{Monopoles acceleration after galaxies formation}\label{sec:astroacc}

If a sizeable number of $\cm$s survived the dilution, as the Universe expanded and cooled down the energy of $\cm$s decreased for the adiabatic energy loss, like any other particle and radiation.
They would have reached a speed $\beta=v/c \sim 10^{-10}$ during the epoch of galaxy formation.
As matter started to condense gravitationally into galaxies, galactic magnetic fields developed through the dynamo mechanism. 

A magnetic field $\mathbf{B}$ acting over a length $\ell$ increases the kinetic energy $K$ of a monopole by a quantity 
\begin{equation}\label{eq:gbl}
K=g\int_{path} \mathbf{B} \cdot d\mathbf{l} \sim gB \ell  \ .
\end{equation}
This equation (in c.g.s. units) is easily derived from the duality relation (\ref{eq:dua}). Using the numerical value of Dirac magnetic charge (\ref{eq:gd}), the strength of the typical galactic magnetic field, $B \sim 3\ 10^{-6}$ G, and a coherent length $\ell \sim 300 \textrm{ pc} = 10^{21}$ cm, a value of $K=10^8$ erg $\simeq 10^{11}$ GeV is obtained.
This means that all $\cm$s with mass $M<10^{11}$ GeV/c$^2$ are expected relativistic on Earth: the energy gain (\ref{eq:gbl}) does not depend on $M$.
For higher masses, non-relativistic velocities $v$ are attained, according to the relation $1/2 M v^2=K$. This justify the expected velocity distribution as a function of the monopole mass in the Particle Data book \cite{pdg}:
\begin{equation}\label{Eq:2.v}
\beta \simeq \left\{
\begin{array}{ll}
1 & M \lesssim 10^{11} \textrm{ GeV/c}^2 \\
10^{-3} \big( { 10^{17}\textrm{ GeV/c}^2\over M }\big)^{1/2} & M \gtrsim 10^{11} \textrm{ GeV/c}^2
\end{array}\right.
\end{equation}
However, this relation can be modified if we assume that $\cm$s can be accelerated by stronger magnetic fields over larger coherence regions. Examples are reported in \cite{wick} and are 
AGN jets ($B\sim 100 \ \mu$G, $\ell\sim 10$ kpc), 
galaxy clusters ($B\sim 30 \ \mu$G, $\ell\sim 100-1000$ kpc), or
extragalactic sheets ($B\sim 1 \ \mu$G, $\ell\sim 30$ Mpc). In these cases, maximum energies of $10^{13}-10^{14}$ GeV can be reached and Eq. (\ref{Eq:2.v}) must be changed accordingly.
Depending on the assumed astrophysical models, we thus expect a flux of relativistic $\cm$s on Earth if their masses are below $10^{11}-10^{14}$ GeV/c$^2$. In particular, IM$\cm$ are expected to be relativistic.

Higher mass $\cm$s (as the GUT ones) would be gravitationally bound to different \textit{systems}. 
Their speeds would be defined by the virial theorem for objects gravitationally bound to the {system} and would be minimally affected by astrophysical magnetic fields. 
Thus, their kinetic energy depends on the mass M, and not on the magnetic charge. 
Thus, $\cm$s trapped locally in the Solar System would have $\beta \simeq 10^{-4}$; $\cm$s bound in the Milky Way would have $\beta \simeq 10^{-3}$ and those bound in the local cluster of galaxies $\beta \sim 10^{-2}$. The $\cm$s extragalactic flux should be isotropic, while the local ones are probably concentrated in the planes of the orbits. 


\subsection{Astrophysical upper bounds}\label{sec:astrobounds}
Different upper limits on the expected flux of cosmic $\cm$s are obtained based on cosmological and astrophysical considerations.
Most of these bounds have to be considered as rough orders of magnitude.
A cosmological bound is obtained by requiring that the present monopole number density $n_M$ is smaller than the critical density (\ref{eq:2.rhoc}), i.e. $\rho_M = n_M M <\rho_c$.
The $\cm$ flux per unit solid angle, $\Phi_M$, is related to $n_M$, to the mass $M$ and to their average speed $v=c\beta$ in the observed frame by:
\begin{equation}\label{eq:2.phi}
\Phi_M= {n_M v \over 4 \pi} < {\rho_c c \over 4 \pi} \cdot {\beta \over M} \simeq 2.5\cdot 10^{-13} { h}^{2} \cdot {\beta \over (M/10^{17} \textrm{ GeV/c}^2) }
 \textrm{ [cm}^{-2} \textrm{s}^{-1} \textrm{sr}^{-1}] \ .
\end{equation}
The limit holds in the hypothesis of poles uniformly distributed in the Universe. 
As the local (i.e. near the Solar System) dark matter density is $\rho_{local}\sim 0.3$ GeV/cm$^{3}$, i.e. about five orders of magnitude higher than $\rho_c$ in (\ref{eq:2.rhoc}), the $local$ flux of $\cm$s could be up to a factor $\sim 10^5$ higher than given by Eq. (\ref{eq:2.phi}). 

A more stringent limit can be derived by the considerations that astrophysical objects do have large-scale magnetic fields. 
For instance in our Galaxy, the magnetic field (probably) originated by the non-uniform rotation of the Galaxy is stretched in the azimuthal direction along the spiral arms. 
The time scale of this generation mechanism is approximately equal to the rotation period of the Galaxy, $\tau_B \sim 10^8$ y. 

Monopoles gain kinetic energy in a magnetic field by reducing the stored magnetic energy density.
An upper bound for the monopole flux can be obtained by requiring that the kinetic energy gained per unit time and per unit volume, $d^2K/dt dV$, is at most equal to the magnetic energy generated by the dynamo effect. 
The rate of kinetic energy gained by magnetic monopoles in a magnetic field $\mathbf{B}$ is
\begin{equation}\label{eq:2.din1}
{d^2K\over dt dV} = \mathbf{J}_M\mathbf{\cdot B} 
\quad \textrm{[erg s}^{-1} \textrm{cm}^{-3}] \ ,
\end{equation}
where $ \mathbf{J}_M = g n_M \mathbf{v}$ represents the magnetic current density and $\mathbf{v}$ is the average pole velocity. 
The magnetic energy density, $\rho_B = B^2/8\pi$, generated by the astrophysical dynamo effect per unit time is
\begin{equation}\label{eq:2.din2}
{d^2E_D\over dt dV} = {\rho_B \over \tau_B} = {B^2 \over 8\pi \tau_B} \quad \textrm{[erg s}^{-1} \textrm{cm}^{-3}] \ .
\end{equation}
Assuming $\mathbf{v}$  and $\mathbf{B}$ parallel over large distances,  the condition $ \mathbf{J}_M \cdot  \mathbf{B} < {B^2 /8\pi \tau_B}$ leads to the upper bound: 
\begin{equation}\label{eq:2.din2bis}
n_M <\frac{B}{8\pi \tau_B g v} \ . 
\end{equation}
As in Eq. \ref{eq:2.phi}, a relation between flux and number density can be obtained
\begin{equation}\label{eq:2.din4}
\Phi_M = {n_M v \over 4 \pi} \lesssim {B \over 32 \pi^2 \tau_B g} \simeq 10^{-16} \ \textrm{ [cm}^{-2} \textrm{s}^{-1} \textrm{sr}^{-1}] \ .
\end{equation}
The numerical values refer to a typical value of the Galactic magnetic field strength, $B \sim 3\ 10^{-6}$ G, and single Dirac charge, $g=g_D$. 
The above condition represents the so-called \textit{Parker bound} \cite{parker}.
Note that the condition in Eq. \ref{eq:2.din4} is always more stringent than Eq. \ref{eq:2.phi} for $M<10^{17}$ GeV/c$^2$ and $\beta>10^{-3}$.

In a more detailed treatment \cite{82T1}, which assumes reasonable choices for the astrophysical parameters and random relation between $\mathbf{v}$ and $\mathbf{B}$, $\cm$s acquire smaller energies. Correspondingly, less energy is removed from the galactic field and the corresponding bound is less restrictive. 
According to Eq. \ref{Eq:2.v}, and defining a critical speed $\beta_c=10^{-3}$,  the upper bounds derived in \cite{82T1} are
\begin{equation}\label{eq:2.din6}
\Phi_M  \lesssim \left\{
\begin{array}{ll}
10^{-15} \quad\quad\quad\quad\quad\quad\quad\quad  \textrm{ [cm}^{-2} \textrm{s}^{-1} \textrm{sr}^{-1}],\quad & M \lesssim 10^{17} \textrm{ GeV/c}^2 \\
10^{-15} \big( { 10^{17}\textrm{ GeV/c}^2\over M }\big) \big( {\beta \over\beta_c} \big)\ \textrm{ [cm}^{-2} \textrm{s}^{-1} \textrm{sr}^{-1}],\quad  & M \gtrsim 10^{17} \textrm{ GeV/c}^2
\end{array}\right.
\end{equation}

Finally, an \textit{extended Parker bound} can be obtained by considering the survival of an early seed field \cite{adams93,lewis00}, yielding the  tighter bound
\begin{equation}\label{eq:2.din7}
\Phi_M \lesssim 1.2\ 10^{-16} \biggl( {M \over 10^{17} \textrm{ GeV/c}^2 }\biggr) \ \
\textrm{ [cm}^{-2} \textrm{s}^{-1} \textrm{sr}^{-1}]  \ .
\end{equation}

\section{Magnetic Monopole Energy Losses}\label{sec:loss}

The interaction mechanisms of $\cm$s with matter are important to understand their stopping power in matter in general and the energy losses in particle detectors in particular. 
Magnetic monopoles are very highly ionising particles, because of their strong magnetic charge (\ref{eq:gd}). A relativistic $\cm$ with minimum magnetic charge $ g_D $ would ionise $\sim 4700$ times more than a minimum ionizing particle, a muon for instance. 

The calculation of the energy loss for $10^{-3}\lesssim \beta \lesssim 10^{-2}$ assumes the materials as a free (degenerate) gas of electrons; $\cm$s thus interact with the conduction electrons of metallic absorbers \cite{ak82}.
For $\beta\lesssim 10^{-3}$, the long-range interaction of the magnetic-dipole moment $\boldsymbol\mu$ of a fermion, or atomic system, or a nucleus, occurs through the static $\cm$ field ${B_M}\propto g/r^2$ \cite{77K1,83B1}.
The corresponding interaction energy is $ W_D = - \boldsymbol\mu \mathbf{\cdot B_M}$. 
For an electron at a distance $r$, it corresponds to $W_D= \hbar^2/4 m_e r^2$ and at a distance equal to the Bohr radius, $r= a_o= 0.53 \times 10^{-8}$ cm, $W_D\simeq 7$ eV, comparable to the atomic binding energies.
Thus, one expects a sizable deformation of an atom when a very-slow monopole passes inside or close to an atomic system.
For a proton at a distance of $r=1$ fm, $W_D \sim 30$ MeV, a value larger than the binding energy of nucleons in nuclei. Thus, one expects deformations of the nucleus when a monopole passes close to it.
A general description of energy losses mechanisms in the range of monopole velocities $\beta > 10^{-5}$ is provided in \cite{patspu}.  
In the following of this section, we concentrate on the energy loss mechanisms connected with the detection of a $\cm$ in a velocity range suitable for neutrino telescopes, i.e. when $\beta>0.1$.

Classical and IM$\cm$s are expected to be relativistic; their energy loss would be sufficiently large to stop a considerable fraction when crossing the Earth.
Magnetic poles with $M>10^{14}$ GeV/c$^2$, and in particular GUT ones, are expected non-relativistic, according to astrophysical acceleration models. 

\subsection{Ionisation energy loss of fast monopoles}
\label{sec:3-eloss-1}
A moving $\cm$ produces an electric field in a plane perpendicular to its trajectory. 
In matter, this field may ionize or excite the nearby atoms or molecules.
The interaction of $\cm$s having velocities $\beta> 0.05$ and $\gamma \le 100$ with the electrons of a material is well understood; for $g=g_D$, a monopole behaves as an equivalent electric charge $(Ze)^2 = g_D^2\beta^2 \sim (68.5\beta)^2$.
The ionization energy loss for electric charges, see \S \textit{32. Passage of particles through matter} of \cite{pdg}, can be adapted for a magnetic charge \cite{ahlen} by replacing $Ze \rightarrow g \beta$. The stopping power for $\cm$s becomes
\begin{equation}\label{eq:3.elos}
{dE\over dx} = {4\pi N_e g^2 e^2 \over m_e c^2} \biggl[ \ln \biggr( {2m_ec^2 \beta^2\gamma^2 \over I}\biggr) +{K \over 2} - {\delta +1 \over 2} - B_m \biggr] \ ,
\end{equation}
where $m_e$ is the electron mass, $I$ the mean ionization potential of the crossed medium, $\delta$ the density effect correction, $K$ a QED correction to consider $\delta$-rays (see \ref{sec:deltach}) , and $B_m$ the Bloch correction (refer to \cite{derk98} for the numerical values for different materials).
$N_e$ is the density of electrons, depending on the Avogadro's number $N_A$ and on the charge and atomic numbers of medium
\begin{equation}\label{eq:Ne}
N_e=\rho N_A \frac{Z}{A}\ \textrm{ [cm}^{-3}]
\end{equation}
The energy loss mechanism (\ref{eq:3.elos}) is implemented as a part of the GEANT package \cite{geant} for the simulation of monopole trajectories in a detector \cite{bauer}.

\begin{figure*}[tb]
\begin{center}
\includegraphics[width=12.0cm]{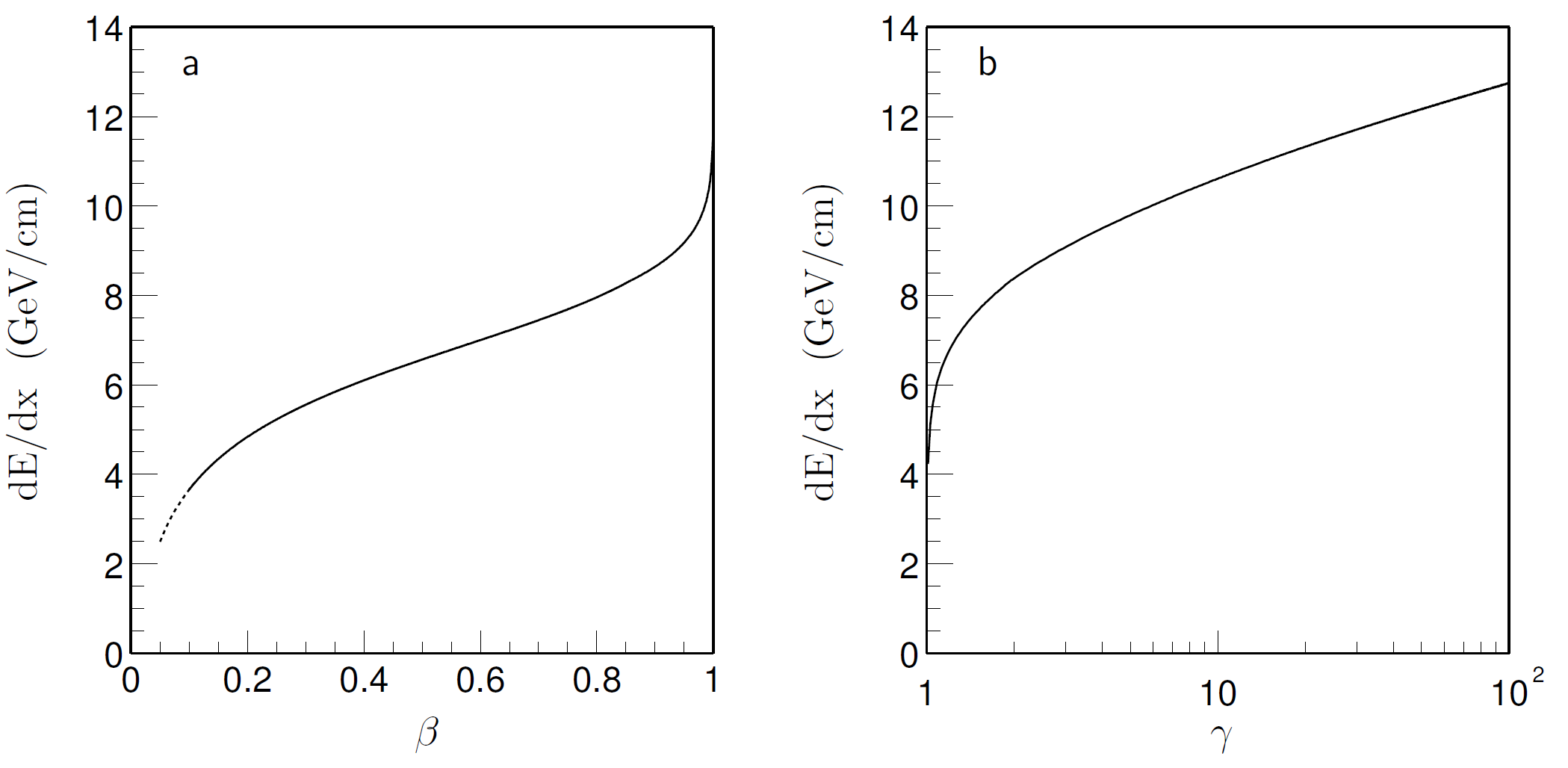}
\end{center}
\caption{\small\label{fig:eloss} a) The energy loss of a monopole with one Dirac charge $g_D$ in the sea water as a function of the monopole velocity $\beta c$ and (b) as a function of the monopole's Lorentz factor $\gamma$. The dashed tail of the curve indicates the limited validity of the underlying assumptions at low  $\beta$. Adapted from \cite{bram}.  }
\end{figure*}

Fig. \ref{fig:eloss} shows the energy loss of a magnetic monopole in water as a function of the monopole $\beta$ (left panel) and for $\gamma<100$ (right panel). The $dE/dx$ is about 3.5 GeV/cm at $\beta= 0.1$ and increases with speed to $\sim$ 13 GeV/cm at $\gamma = 100$. At higher energies, for $100<\gamma<10^4$, the energy loss can be safely extrapolated according to Eq. \ref{eq:3.elos}, i.e., with a logarithmic increases, although some approximations are not valid anymore.

For $\cm$ energies such that $\gamma>10^4$, stochastic processes such as bremsstrahlung, pair production and hadron production dominate the energy loss. 
Photonuclear processes yield the largest contribution in materials with moderate values of atomic mass (such as rock or water), with a smaller contribution from pair production. 
In contrast, pair production is the dominant process in materials with high atomic number $Z$.
The contribution from bremsstrahlung is small and is usually neglected.
Due to the value of the magnetic charge (\ref{eq:gd}), the equivalent of the electromagnetic coupling constant for monopoles, $\alpha_m$, is large and thus perturbative calculations are not allowed. This imply very large theoretical uncertainties, which must be considered in the computation of transition probabilities.
The energy loss per unit path length in this energy regime is thus dominated by single (or few) collisions with large energy transfer, leading to large fluctuations on the stopping power and range of the particle. 

\subsection{Cherenkov radiation\label{sec:cherenkov} }

A $\cm$ induces Cherenkov radiation when it passes through a medium of refractive index $n$ with a velocity larger than the phase velocity, $c/n$, of light in the medium ($n\simeq 1.35$ for seawater, and $n\simeq 1.31$ for ice).
The minimum monopole velocity for Cherenkov light emission to occur in water (ice) is thus about 0.74c (0.76c). This threshold velocity is referred to as the \textit{Cherenkov limit}.
The Cherenkov photons are emitted at the characteristic angle $\theta_C$ with respect to the direction of the $\cm$ such that:
\begin{equation}\label{eq:chea}
\cos\theta_C = \frac{1}{\beta n}
\end{equation}
The number of Cherenkov photons, $N_\gamma $, emitted by a $\cm$ with magnetic charge $g$ per unit path length $dx$ and unit photon wavelength interval $d\lambda$ is described by the Frank-Tamm formula 
for a particle with effective charge $Ze\rightarrow gn$ \cite{tomp}:
\begin{equation}\label{eq:tamm}
\frac{d^2N_\gamma}{dxd\lambda}=\frac{2\pi\alpha}{\lambda^2}
\biggl(\frac{gn}{e}\biggr)^2
\biggl(1-\frac{1}{\beta^2 n^2}\biggr) \ ,
\end{equation}
where $\alpha \simeq 1/137$ is the electromagnetic coupling constant. The number of photons provided by Eq. (\ref{eq:tamm}) is a factor $(gn/e)^2\simeq 8200$ more than that emitted by a particle with electric charge $e$ at the same velocity. A relativistic $\cm$ with $g=g_D$ emits about $3\times 10^6$ photons with wavelengths between $\lambda=300 \div 600$ nm per centimetre path length. 
The corresponding energy loss of the monopole is about 3 MeV/cm, which is only few per mill of its total energy loss (see Fig. \ref{fig:eloss}). The mentioned wavelength interval of the photons corresponds to the sensitive range of typical photo-multiplier tubes used by neutrino telescopes \cite{Chiarusi}. 

\subsection{Cherenkov radiation from $\delta$-rays\label{sec:deltach} }

Part of the ionization energy loss, connected to the term $K$ in (\ref{eq:3.elos}), is transferred to electrons in collisions large enough to knock the electrons out of their atomic orbits. These electrons are referred to as \textit{knock-on electrons} or $\delta$-rays, and have enough kinetic energy to travel short distances through the medium. 
Like any electrically charged particle, they emit Cherenkov light when their velocity is $v_e>c/n$\footnote{In this section, subscript $e$ and $m$ are used to distinguish parameters related to the speed of $\delta$-rays and monopoles, respectively.}. 
The distribution of $\delta$-rays produced by a monopole can be derived from the similar distribution produced by a heavy electric charge, as in \S 32 of \cite{pdg}. 
The knock-on process corresponds to a Coulomb collision of a heavy incident particle with electric charge $ze$ and velocity $\beta c$ with a free stationary electron. Neglecting particle spins, the cross section, per electron energy interval $dT_e$, is: 
\begin{equation}\label{eq:ste}
\frac{d\sigma}{dT_e}=\frac{2\pi z^2 e^4}{m_e c^2 \beta^2 T_e^2} \ .
\end{equation}
 
This expression derives from the Rutherford scattering formula, and is valid for close collisions and when the atomic electrons can be considered as free. This corresponds to the condition that $T_e$ is much larger than the mean ionization energy of the medium, $I$, which holds for relativistic $\delta$-rays and $I \sim 10$ eV. 
The formula for a $\cm$ of charge $g$ and velocity $\beta_m c$ is obtained by replacing $ze\rightarrow g\beta_m$.
For relativistic velocities, and considering the effect of the electron spin in the collision, Eq. (\ref{eq:ste}) is modified to
\begin{equation}\label{eq:mott}
\biggl( \frac{d\sigma}{dT_e}\biggr)_M 
=\frac{2\pi g^2 e^2}{m_e c^2 T_e^2} 
\biggl(1-\beta_m^2 \frac{T_e}{T_e^{m}}    \biggr) \ ,
\end{equation}
where $ T_e^{m}$ is the upper limit on the energy that can be transferred to an atomic electron in a single collision. The relation is equivalent of the so-called Mott cross section for charged particles. 

As the masses of monopoles are much larger than the electron mass, $m_e$, the maximum energy transfer derived from the Coulomb scattering is: 
\begin{equation}\label{eq:tmaxI}
T_e^{m} = 2 m_e c^2 \beta_m^2 \gamma_m^2
\end{equation}
From this relation, it follows that a monopole must have a velocity $\beta_m c\gtrsim0.44$ $(0.47)$ to be able to produce $\delta$-rays energetic enough to emit Cherenkov light in water (ice).
We derive this important result for water. To produce Cherenkov light, the velocity of $\delta$-ray must be above the Cherenkov threshold, $\beta_e >0.74$, corresponding to $\gamma_e= 1.49$. 
This corresponds to a kinetic energy of the electron of $\sim 0.25$ MeV.
Thus, the maximum transferred total energy (adding the electron rest mass) of the $\delta$-ray must satisfy the condition: 
\begin{equation}\label{eq:limI}
2 m_e c^2 \beta_m^2 \gamma_m^2 +  m_e c^2 > m_e c^2 \gamma_e 
\end{equation}
or
\begin{equation}\label{eq:limII}
 2 \beta_m^2 \gamma_m^2 > (\gamma_e -1) 
\end{equation}
which holds if $\beta_m >0.44$.
Similar values can be derived for ice, using the different Cherenkov threshold, $\beta_e >0.76$.

\begin{figure*}[tb]
\begin{center}
\includegraphics[width=12.0cm]{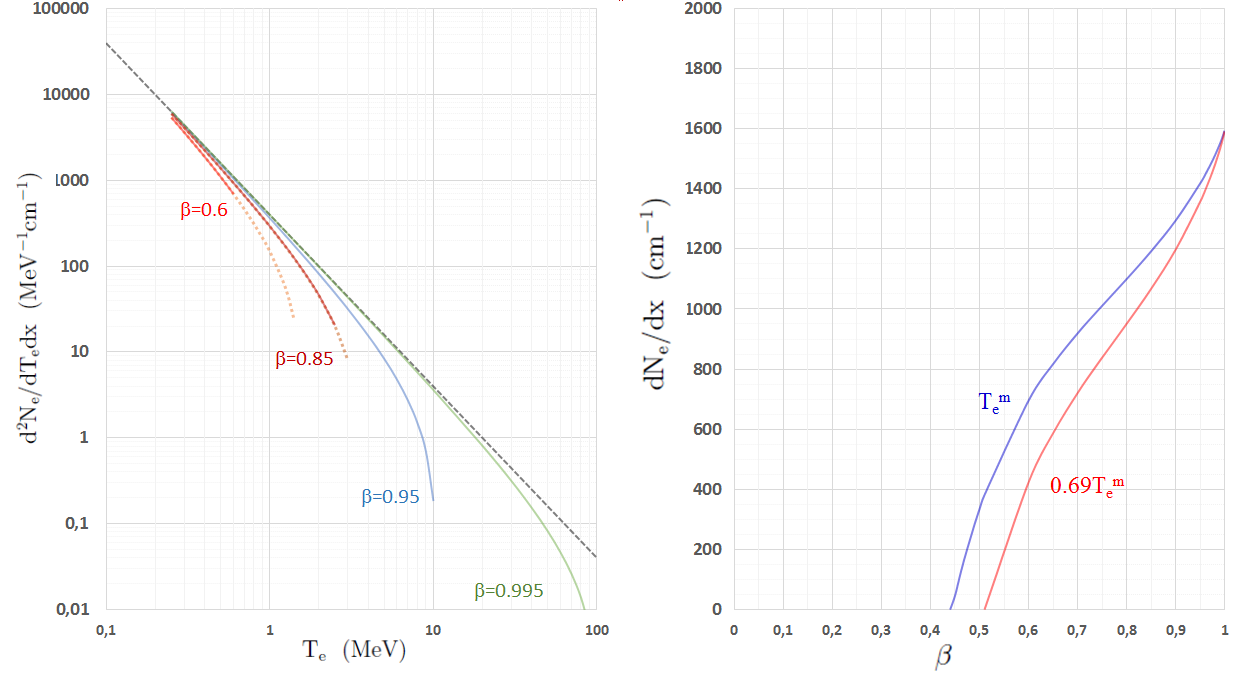}
\end{center}
\caption{\small\label{fig:deltaspectrum} 
Left: The differential distribution of $\delta$-rays with kinetic energies $>0.25$ MeV produced by a monopole with one Dirac charge, $g_D$, passing through one cm of water. The distribution is shown for monopole velocities 0.60c, 0.85c, 0.95c and 0.995c. This last value corresponds to $\gamma\simeq 10$. Dashed lines corresponds to energies above the kinematic thresholds. The black dashed line indicates a spectrum that is proportional to $1/T_e^2$.
Right: The total number of $\delta$-rays with kinetic energies above 0.25 MeV produced in water per centimetre path length by a $g_D$ monopole, as a function of the monopole $\beta$. Blue: maximum energy given by Eq. (\ref{eq:tmaxI}); Red: maximum energy given by Eq. (\ref{eq:tmaxII}).}
\end{figure*}

The differential number of $\delta$-rays per unit path length and unit kinetic energy interval is obtained by multiplying Eq. (\ref{eq:mott}) (which has units: cm$^2$ MeV$^{-1}$) to the electron number density (\ref{eq:Ne}). 
For water or ice with density $\rho \simeq 1$ g/cm$^3$, A=18, Z=10 we have $N_e\simeq 3\times 10^{23}$. Thus, the differential distribution of $\delta$-rays per unit path length for $g=g_D$ is given by:
\begin{equation}\label{eq:mottI}
\biggl( \frac{d^2 N_e}{dT_e dx}\biggr)_M 
=\frac{2\pi g^2 e^2 N_e}{m_e c^2 T_e^2} 
\biggl(1-\beta_m^2 \frac{T_e}{T_e^{m}} \biggr) 
\simeq \frac{400}{ T_e^2} \biggl(1-\beta_m^2 \frac{T_e}{T_e^{m}}\biggr)  \ ,
\end{equation}
where in the last equality we have inserted the numerical values.
Fig. \ref{fig:deltaspectrum} (left) shows the differential distribution of $\delta$-rays obtained for four different values of the monopole velocity, yielding maximum transferred energy as in Eq. (\ref{eq:tmaxI}). 
The total number of $\delta$-rays with energies above $T_0=0.25$ MeV that are produced per cm in water is determined by integrating Eq. (\ref{eq:mottI}) starting from $T_0$. The result is shown in Fig. \ref{fig:deltaspectrum} (right) as a function of the monopole velocity (blue line). The number of $\delta$-rays produced per centimetre is zero at the threshold $\beta_m=0.44$, and increases to about 1600 to $\beta=1$.

A more conservative value of the maximum transferred energy is derived in \cite{ah88}, where 
\begin{equation}\label{eq:tmaxII}
 T_e^{max} = 0.69\cdot T_e^{m}
\end{equation}
According to this condition, it follows that a $\cm$ must have $\beta c \gtrsim 0.51$ $(0.53)$ to be able to produce $\delta$-rays energetic enough to emit Cherenkov light in water (ice). In addition, also the number of $\delta$-rays produced by a monopole of a given velocity is reduced, as shown by the red curve in Fig. \ref{fig:deltaspectrum} (right).

A different quantum-mechanical correction for the electron-$\cm$ spin coupling with respect to the Mott cross-section (\ref{eq:mott}) has been considered in \cite{77K1}. Here, the helicity-flip and helicity-nonflip scattering amplitudes of a Dirac particle with spin 1/2 and charge $ze$ by a fixed $\cm$ field were accounted for with a $F(T_e)$  form-factor term instead of the $(1-\beta_m^2 \frac{T_e}{T_e^{m}})$ term in   (\ref{eq:mott}). 
The corresponding KYG (from the three authors of the paper) magnetic monopole cross-section gives a harder $\delta$-ray spectrum than the Mott cross section both for the standard (\ref{eq:tmaxI}) or conservative (\ref{eq:tmaxII}) values of the maximum transferred energy to the electron. 

\begin{figure*}[tb]
\begin{center}
\includegraphics[width=12.0cm]{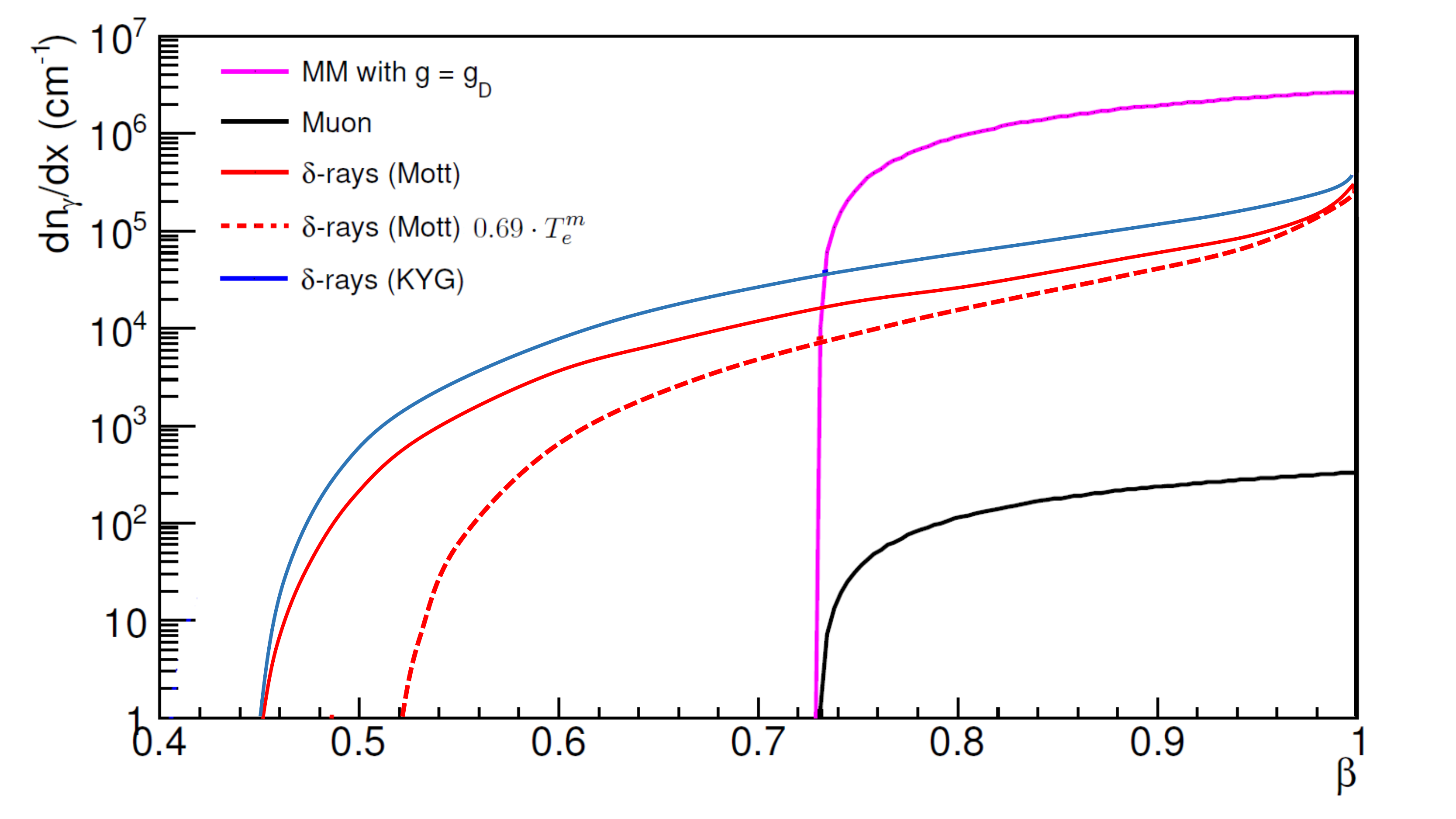}
\end{center}
\caption{\small\label{fig:drg} Light yield in water of a $\cm$ with magnetic charge $g_D$ due to the Cherenkov effect (cyan) and due to the $\delta$-rays production. The spectrum derived with $T_e^{m} = 2 m_e c^2 \beta_m^2 \gamma_m^2$ from the Mott cross-section is in red and from the KYG cross-section in blue; that for the Mott cross-section and $0.69 T_e^{m}$ with dashed red line (see text). For comparison, the light yield for a muon is also shown (black). }
\end{figure*}

The number of Cherenkov photons, $N_\gamma$, emitted by a $\delta$-ray in the sea water per unit path length $dx$ and unit wavelength interval is derived from the Frank-Tamm formula (\ref{eq:tamm}). In the wavelengths range between 300 and 600 nm, it amounts to about:
\begin{equation}\label{eq:fw}
\frac{dN_\gamma}{dx} \simeq 760 \biggl( 1-\frac{1}{\beta_e^2 n^2} \biggr) \ \textrm{ cm}^{-1}
\ ,
\end{equation}
The total number of photons emitted by a $\delta$-ray thus depends on the electron's velocity and on its path length before its kinetic energy drops below $T_0=0.25$ MeV.
In turn, the $\delta$-ray's path length and velocity depend on its initial kinetic energy and energy loss in the water (due to the involved energies, only the ionisation energy loss must be considered).
This can be obtained using full Monte Carlo simulations, or following analytic approximations, as presented in \cite{bram}.
According to this numerical approximation, the total number of Cherenkov photons, $n_\gamma$, that is emitted by all $\delta$-rays that are produced per unit path length of a monopole is shown in Fig. \ref{fig:drg} as a function of the monopole velocity.
The result is very similar to that obtained by the ANTARES collaboration in water \cite{antares2} and by the IceCube collaboration in ice \cite{ic2}.

\subsection{H$_2$0 molecules luminescence \label{sec:3-lumi} }
In addition to directly induced Cherenkov radiation, or that induced by secondary $\delta$-rays, a third process which may be considered for neutrino telescopes in water or ice \textit{luminescence}.

The {luminescence} is the fraction of the energy loss (\ref{eq:3.elos}) that goes in excitations of atomic/molecular levels and resulting in visible light.
The observables of luminescence, such as the wavelength spectrum and decay times, are highly dependent on the properties of the medium, in particular, temperature and purity. 
Contrarily to the Cherenkov emission of the polarized medium, which is a very fast process, the luminescence decay times are slower. Their knowledge is of particular relevance for the discrimination of a possible signal from the background. 

Concerning ice and water, the results for the light yield of luminescence of different measurements vary by large factors: a summary of the status of observations is in \cite{poll}.
At present, the contribution of this effect for the detection of $\cm$s is not completely understood and not used in published results (see \S \ref{sec:searchesNT}). 
It is expected that the luminescence light production by a $\cm$ with $\beta_m>0.5$ is about one order of magnitude less than the light produced by secondary $\delta$-rays.  However, it could provide a measurable amount of photons for velocities down to $\beta_m\sim 0.1$, or even lower velocities. 
Magnetic monopoles with speed between 0.1c to and 0.5c inducing luminescence light could produce a signature that can be recognized by current data acquisition systems and trigger logics of neutrino telescopes. Slower and fainter events probably require an upgrade in the data acquisition and trigger logic.

\subsection{Range of magnetic monopoles\label{sec:3-range} }
Depending on their stopping power, $\cm$s can cross large amounts of materials, as the Earth diameter, a depth of many km of water equivalent, or the atmospheric layer.
The stopping power depends on the $\cm$ energy loss, on the mass $M$ and initial velocity.

\begin{figure*}[tb]
\begin{center}
\includegraphics[width=9.0cm]{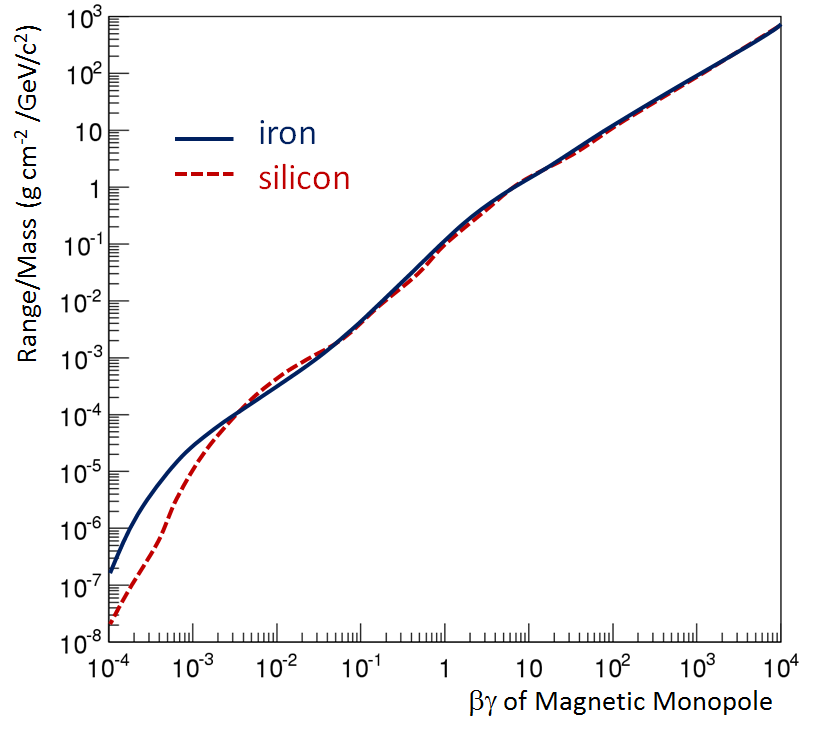}
\end{center}
\caption{\small\label{fig:rm} 
The ratio range/mass for $\cm$s with $g_D$ magnetic charge in iron and silicon as a function of monopole $\beta\gamma=p/M$. The range is computed from the stopping power given in \cite{derk98} and it is defined as the thickness of material to slow down $\cm$s to $\beta = 10^{-5}$. }
\end{figure*}

The range $R$ can be computed by integrating the stopping power:
\begin{equation}\label{eq:3.range}
R=\int_{E_{min}}^{E_0} {dE\over dE/dx} \ ,
\end{equation}
where $E_0$ is the $\cm$ initial kinetic energy, and $E_{min}$ the kinetic energy when it can be considered as stopped (values as low as $\beta \simeq 10^{-5}$ are enough).
 
Fig. \ref{fig:rm} shows the ratio, $R/M$, between range (in g/cm$^{2}$) to mass (GeV/c$^{2}$), as a function of the initial $\beta\gamma=p/M$ of the $\cm$; the quantity $p$ is its momentum \cite{burdin}. The stopping power used in the computation is that obtained in Si and Fe in \cite{derk98}.  
Since the range depends on the kinetic energy $K$, the $R/M$ ratio is independent of the mass.

From Fig. \ref{fig:rm} for, e.g., a $\beta=0.7$ monopole ($\beta\gamma=1$) it corresponds $R/M \sim 0.1$ g cm$^{-2}$/GeV/c$^{2}$.
Thus, to cross the Earth diameter ($R \sim 9\times 10^9$ g/cm$^{2}$), the $\cm$ mass must be $M \gtrsim 10^{11}$ GeV/c$^{2}$; to cross 3000 m.w.e. (a typical depth of underground detectors), $M>3\ 10^{6}$ GeV/c$^{2}$; to cross the atmosphere (1000 g/cm$^{2}$), $M\simeq 10^4$ GeV/c$^{2}$.
Similarly, to cross the Earth, a $\cm$ with $\gamma=10^4$ should have mass larger than $10^7$ GeV/c$^{2}$.

The above estimates have an effect on the value of $\cm$ masses that can be investigated by neutrino telescopes. According to the discussion of \S \ref{sec:astroacc}, monopoles can gain kinetic energies from $K\simeq 10^{11}$ GeV (in magnetic fields of normal galaxies, as our own Milky Way) to $K\simeq 10^{14}$ GeV in galaxy clusters.
The kinetic energy, $K\simeq 10^{11}$ GeV, gained by galactic magnetic fields is just sufficient to allow monopoles with $M\lesssim 10^{11}$ GeV/c$^{2}$ to be relativistic and cross the Earth diameter. 
For very low monopole masses, when $\gamma_m= K/ Mc^2>10^4$, i.e. when $M<10^{7}$ GeV/c$^2$, the stochastic processes (such as pair production and hadron production) dominate the energy loss mechanism, and $\cm$s are absorbed by crossing the Earth.

\begin{figure*}[tb]
\begin{center}
\includegraphics[width=9.0cm]{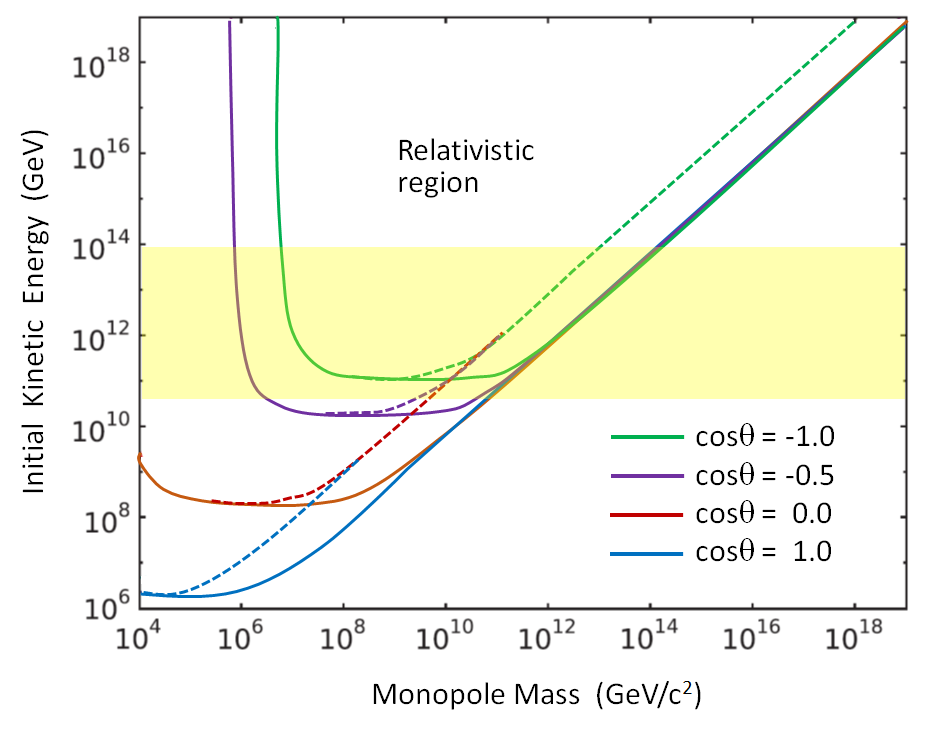}
\end{center}
\caption{\small\label{fig:KM} 
Parameter space of $\cm$s at the Earth's surface required to reach a neutrino telescope with relativistic speeds. Dotted is for a speed threshold of $\beta= 0.995$, while the solid is for $\beta=0.8$.
The four different zenith angles correspond to $\cm$s crossing different amounts of material, from the Earth diameter ($\cos\theta=-1.0$) to 2000 m.w.e. ($\cos\theta=+1.0$). The yellow band corresponds to the kinetic energies gained by $\cm$s from astrophysical magnetic fields. Adapted from \cite{ic1} .}
\end{figure*}
In general, $\cm$s with masses such that $Mc^2<K$ are relativistic.
Only those with masses $Mc^2\gtrsim K/10^4$ are in the energy loss regime described by Eq.  (\ref{eq:3.elos}), and can cross the Earth diameter to be detected as up going particles in an underwater/ice detector. 
For $Mc^2< K/10^4$, the energy loss is much larger and predictions are not straightforward. 
Fig. \ref{fig:KM} shows the region of parameters $K$ (kinetic energy) and $\cm$ mass yielding a visible $\cm$ in a neutrino telescope. Depending on selection criteria to remove the background, sometime only upgoing events (corresponding to $\cos\theta <0$ values, where $\theta$ is the zenith angle) are selected. The yellow region shows the range of kinetic energy attainable by $\cm$s through acceleration in astrophysical magnetic fields (see \S \ref{sec:astroacc}).  


\subsection{Monopoles inducing catalysis of nucleon decay}
\label{sec:cata}

\begin{figure*}[tb]
\begin{center}
\includegraphics[width=12.0cm]{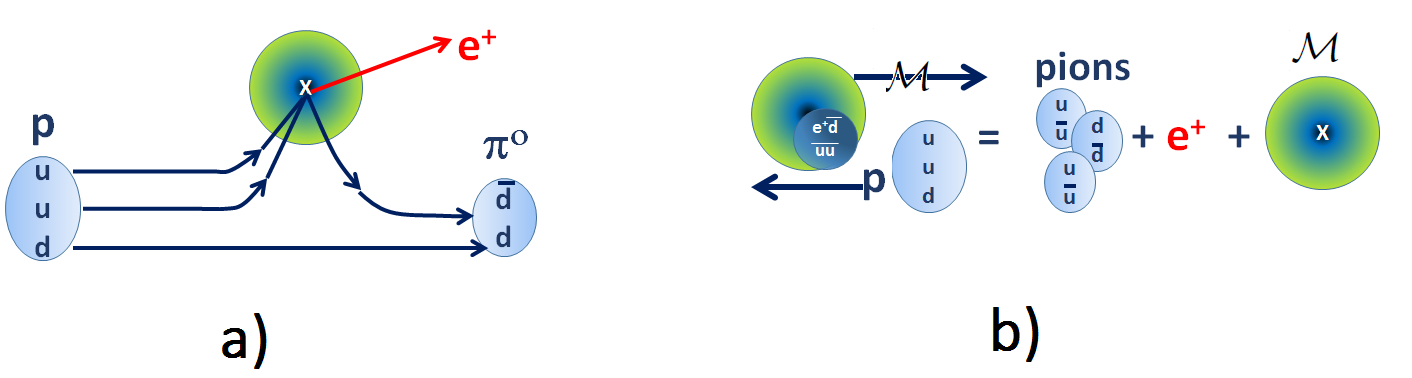}
\end{center}
\caption{\small\label{fig:inner} 
Illustrations of the monopole catalysis of proton decay through the reaction $ p + \cm \rightarrow \cm + e^+ +\pi^0$ (a). 
The effect of the presence of a condensate of 4-fermions $\overline u \overline u \overline d e^+$ able to induce the proton decay is shown in (b).}
\end{figure*}

Given the inner structure of a GUT $\cm$ shown in Fig. \ref{fig:gutM}, it was hypothesized that it could catalyse baryon number violating processes, such as 
\begin{equation}\label{eq:cata}
p + \cm \rightarrow \cm + e^+ + \textrm{ mesons}
\end{equation}
The cross section $\sigma_0$ of this process would be very small, of the order of the geometrical dimension of the $\cm$ core ($\sim 10^{-58}$ cm$^2$).
If the interaction is independent on the presence of $X,Y$ bosons in the monopole core, the catalysis cross section $\sigma_0$ would be comparable to that of ordinary strong interactions, through the so-called Rubakov-Callan mechanisms \cite{81R1,82C4}.
The catalysis reaction on the $\cm$ core can be imagined pictorially as shown in the left part of Fig. \ref{fig:inner}, while the effect of the presence of a condensate of 4-fermions $\overline u \overline u \overline d e^+$ is illustrated in the right part.

The effective catalysis cross section $\sigma_{cat}$ depends also on the $\cm$ speed, as $\sigma_{cat}= (\sigma_0 /\beta) \cdot F(\beta)$.
The form-factor $F(\beta)$ takes into account the angular momentum of the $\cm$-nucleus system and becomes relevant for speeds below a threshold $\beta_0$, which depends on the nucleus \cite{cata_b}. 
For speeds above $\beta_0$, $F(\beta)=1$. Current estimates for the catalysis cross sections are of the order of $10^{-27}$ cm$^2$ to $10^{-21}$ cm$^2$ \cite{nath}.
Due to their different inner structure, the catalysis process is not possible for IM$\cm$s. 
Therefore, the sensitivity of dedicated searches described in \S \ref{sec:searchescata} are constrained to GUT $\cm$s.

Finally, we should note that positively charged dyons\footnote{A dyon is a particle carrying both electric and magnetic charges.}, or ($\cm+p$) states, should not catalyse proton decay with large rates because of the electrostatic repulsion between the proton and the dyon. Negatively charged dyons would instead have large effective cross sections.

\section{Searches for Magnetic Monopoles}
\label{sec:searches}

Searches for $\cm$s have been performed in cosmic rays, at accelerators, and for monopoles trapped in matter. 
No confirmed observation of particles possessing magnetic charge exists. 
Search strategies for $\cm$s depend on their expected interactions as they pass through a detector. 
They are based on different techniques as, e.g., the induction method, exploiting the electromagnetic interaction of the $\cm$ with the quantum state of a superconducting ring, or the excitation/ionization energy losses in different detectors.
Dedicated analyses are required to search for $\cm$s inducing the catalysis of proton decay.
Here, we just summarize the most relevant results and we refer to \cite{patspu} for a more general review.
The results obtained by neutrino telescopes are described in \S \ref{sec:searchesNT}.

\vskip 0.2cm
\noindent\textbf{The induction technique in superconductive coils.}
The most direct detection method consists in the use of superconducting coils coupled to a SQUID (Superconducting Quantum Interferometer Device). 
A moving $\cm$ of charge $g_D$ induces in a superconducting ring an electromotive force and a current change $(\Delta i)$. For a coil with $N$ turns and inductance $L$ the current change is $\Delta i = 4\pi N n g_D/L = 2\Delta i_o$, where $i_o$ is the current change corresponding to a change of one unit of the flux quantum of superconductivity. 
SQUID have been used in the early searches for $\cm$s in the cosmic radiation, with the limitation of very small acceptances, well below that needed to reach the Parker bound. The technique is still widely used in searches for $\cm$s trapped in matter, as described in \S 5 of \cite{burdin}.

\vskip 0.2cm
\noindent\textbf{Searches at colliders.}
The possible production of Dirac $\cm$s has been investigated at $e^{+}e^{-} $, $e^{+}p$, $p\overline{p}$ and $pp$ colliders, mostly using scintillation counters, wire chambers and nuclear track detectors \cite{fair07}. 
Searches for monopoles trapped in the material surrounding the collision regions were also made.
The searches have set upper limits on $\cm$ production cross sections for masses below the few TeV scale. These limits, based on assumptions of the production processes of monopole-antimonopole pairs, are usually model-dependent.

\vskip 0.2cm
\noindent\textbf{Direct searches in the cosmic radiation.}
Direct searches for $\cm$s in cosmic rays refer to experiments in which the passage of particles is recorded by detectors under controlled conditions \cite{gps}.
For these searches, the background is mainly due to muons of the cosmic radiation \cite{cecco} and natural radioactivity. 
Usually, large detectors have been installed in underground laboratories. 
The minimum mass of $\cm$s to reach these detectors from above (at a given velocity) depends on the overburden of the experiment \cite{SLIMnuclea}.
Each search has to theoretically proof that the used experimental method is sensible to the passage of particles with a magnetic charge in a given mass $M$ interval and $\beta$ range.

\begin{figure*}[tb]
\begin{center}
\includegraphics[width=11.5cm]{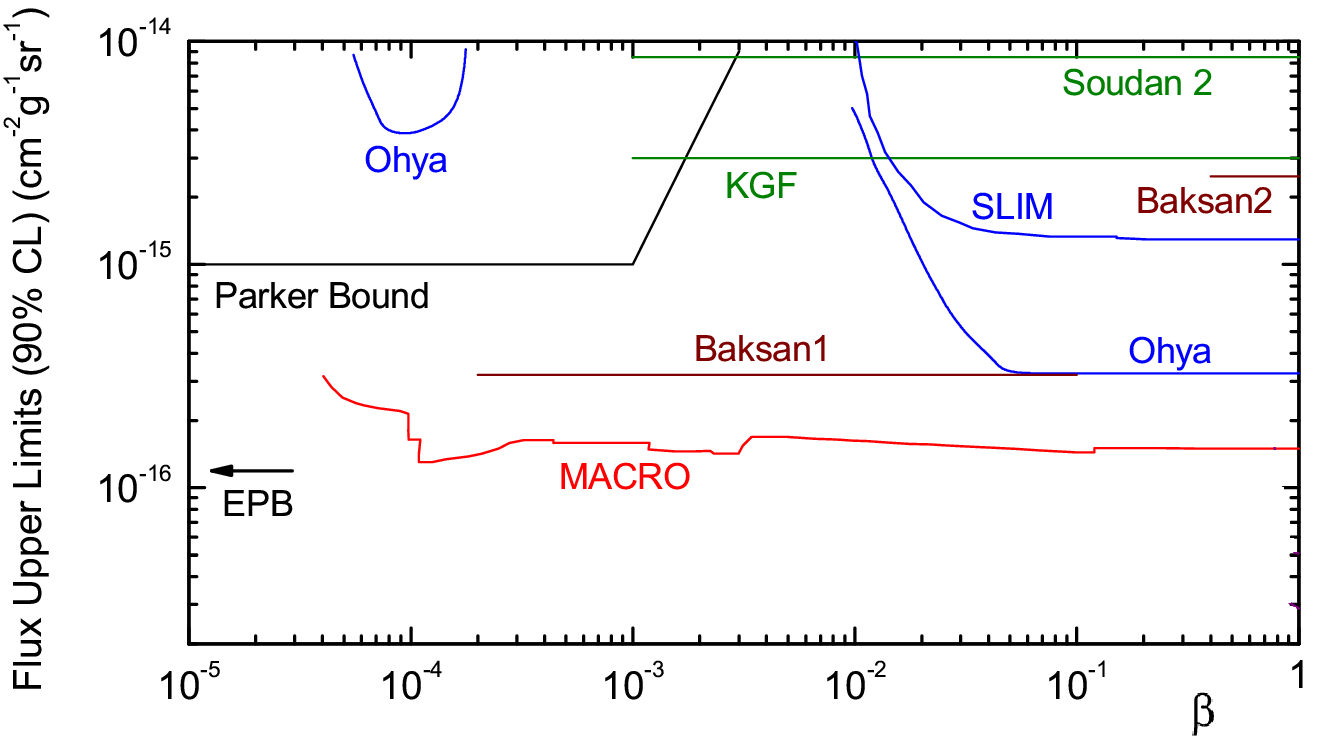}
\end{center}
\caption{\small\label{fig:under} The 90\% CL upper limits vs $\beta$ for a flux of cosmic GUT monopoles with magnetic charge $g=g_{D}$.
The Parker bound refers to Eq. \ref{eq:2.din6}, while the extended Parker bound (EPB) refers to Eq. \ref{eq:2.din7}. For experiments: Ohya \cite{ohya}, Baksan,
Soudan 2 \cite{sou2}, Kolar Gold Field (KGF) \cite{kgf}, SLIM \cite{SLIMmono} and MACRO \cite{02A3}.}
\end{figure*}

Different searches for $\cm$ using large detectors were performed starting from the 1980s \cite{gg84}. 
Fig. \ref{fig:under} shows the 90\% CL flux upper limits versus $\beta$ for GUT $\cm$s with $g=g_{D}$ from the searches with the largest experiments, yielding the better upper limits. 

The most stringent flux upper limits on supermassive $\cm$s in the widest $\beta$ range obtained using different experimental techniques were set by the MACRO experiment \cite{02A4}, who represents a benchmark for magnetic monopole searches.
To accomplish this, the searches were done with an apparatus consisted of three independent sub-detectors: liquid scintillation counters \cite{MAsci,MAall1}, limited streamer tubes \cite{MAst}, each of them with dedicated and independent hardware for $\cm$ searches, and nuclear track detectors. 

\section{Strange quark matter}
\label{sec:sqm}

In 1984 Witten formulated the hypothesis \cite{witten} that strange quark matter (SQM) composed of comparable amounts of $u,\ d$ and $s$ quarks might be the ground state of hadronic matter. 
A system in which an $s$ quark (mass of $\sim 100$ MeV/c$^2$) is replacing either an $u$ or $d$ quark (both with masses $\lesssim 5$ MeV/c$^2$) seems energetically unfavourable. 
However, the energy connected to the sort of \textit{chemical potential} due to the Pauli Exclusion Principle in bulk quark matter could be significantly more than the $s$ mass. Thus, once formed, SQM with three quark flavours in co-existence would be energetically more favourable than ordinary matter and stable. 

The structure of SQM can be described in terms of a nuclear bag model \cite{bag} and only colour singlets can be formed with integer electric charges. 
In order to equilibrate the chemical potential of the quark species, SQM should have a number of $s$ quarks slightly lower than the number of $u$ or $d$ quarks. Thus, the SQM bag have a positive electric charge, which is predicted to be significantly lower than in the case of nuclei, approximately $Z \sim 0.3\ A^{2/3}$. They could be stable for baryon numbers $A$ ranging from a few to $10^{57}$, beyond which a collapse into black holes occurs.

The overall neutrality of SQM is ensured by an electron cloud that surrounds it, forming a sort of atom. The number of electrons are $N_e \simeq (N_d - N_s)/3$, where $N_d$, $N_s$ and $N_e$ are the numbers of quarks $d, s$ and electrons, respectively, assuming $N_d = N_u$.


As magnetic monopoles, SQM could have been produced in a phase transition in the early universe. A tentative mechanism in which a large fraction of the mass and baryon number has survived in the form of nuggets of SQM with radii between 0.1 to 10 cm was offered by Witten \cite{witten}. These objects appeared in the QCD transition from a quasi-free quark-gluon plasma to a colder state where quark are confined. SQM is nonluminous, and did not participate in primordial nucleosynthesis. 
Primordial bags of SQM may constitute part of the dark missing mass of the Universe and of the Milky Way; they thus might be found in the cosmic radiation reaching the Earth. 

Usually, small ($A < 10^7$) SQM systems are called \textit{strangelets}.
The searches of strangelets are almost outside the range of neutrino telescopes: a strangelet is searched for as an event with anomalous charge-to-mass ratio in the cosmic radiation using balloon or space-borne spectrometers. 

The term \textit{nuclearites} is used to design higher mass ($A > 10^7$) objects. Nuclearites are electrically neutral atom-like systems, as they would be expected to possess a electron cloud around the core. For $A > 10^{15}$, electrons would be largely contained within the bag of nuclear matter. 
Nuclearites are generally assumed to be bound to astrophysical objects (the Galaxy, the galaxy cluster,..) and to have speed determined by the virial theorem. In the case of nuclearites bound to the Milky Way, $v\sim 10^{-3} c$. 

The main energy loss mechanism for nuclearites passing through matter is elastic or quasi-elastic atomic collisions \cite{deru}.
The energy loss rate is
\begin{equation}\label{eq:dedxnuclea}
\frac{dE}{dx}= -\sigma \rho v^2 \ ,
\end{equation}
where $\sigma$ is the nuclearite cross section, $v$ its velocity and
$\rho$ the mass density of the traversed medium.
In the case of a nuclearite larger than an atom ($R_0 \gtrsim 10^{-8}$ cm), its cross-sectional area is simply $\sigma \simeq \pi R_0^2$. On the other hand, the effective area of smaller objects is controlled by its {}``electronic atmosphere'' which is never smaller than $\sim 1$ \AA. 
The SQM density corresponds to $\rho_N\simeq 3.5 \times 10^{14}$ g/cm$^3$ (somewhat larger than that of ordinary nuclei).
Thus, nuclearites with masses $M < 8.4 \times 10^{14}$ GeV/c$^2$, the  collisions with ordinary matter are governed by their electronic clouds, yielding $\sigma \simeq \pi 10^{-16}$ cm$^2$.
For nuclearites with radii larger than 1 \AA, i.e. with masses $M \ge 8.4 \times 10^{14}$ GeV/c$^2$, the cross section may be approximated as \cite{deru}:
\begin{equation}\label{eq:sigmanuclea}
\sigma\simeq \pi \times \biggl( \frac{3M}{4\pi \rho_N} \biggr)^{2/3}   \ .
\end{equation}

Transparent media (e.g., scintillators and water) have been used in nuclearites searches. 
Concerning water (or ice), nuclearites do not produce Cherenkov light, as they are non-relativistic objects. They would give rise to a thermal shock through collisions with the atoms of water. The temperature of the medium surrounding the nuclearite path length rises to order of keV. Thus, a hot plasma is formed that moves outward as a shock wave, emitting blackbody radiation and producing a large number of photons in the visible band. 
A detailed description of the luminous efficiency is in \cite{deru}. The authors estimate that in pure water, a fraction of about $3\times 10^{-5}$ of the total energy loss is provided in form of visible light. 

\begin{figure*}[tb]
\begin{center}
\includegraphics[width=11.5cm]{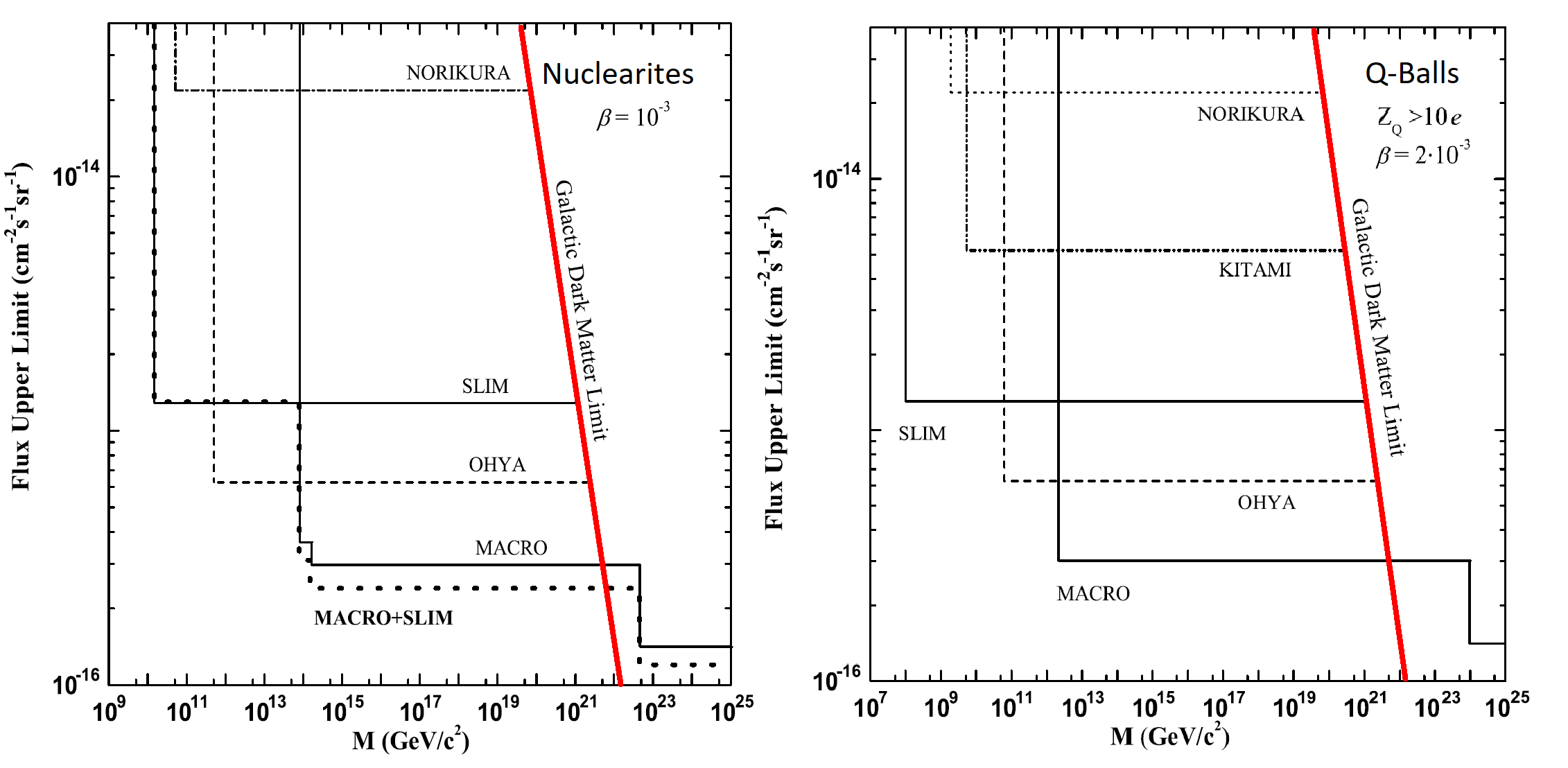}
\end{center}
\caption{\small\label{fig:n+q} Left: 90\% C.L. flux upper limits vs mass for intermediate and high mass nuclearites with $\beta = 10^{-3}$ obtained from various searches with nuclear track detectors. 
Right: 90\% C.L. upper limits of charged Q-balls with $Z_Q > 10 e$ versus mass obtained by various experiments. The red line shows the upper limit provided by the value of the local dark matter density, $\rho_{local}\sim 0.3$ GeV/cm$^{3}$.
See \cite{SLIMnuclea} for references and experimental methods of quoted experiments.}
\end{figure*}

Any search for nuclearites has an acceptance that depends on nuclearite mass; only nuclearites with sufficiently large mass ($>6 \times 10^{22}$ GeV/c$^2$) can traverse the Earth at typical galactic velocities \cite{deru}.
A summary of the detection techniques and experimental results can be found in \cite{MACROnuclea,SLIMnuclea} and in the left plot of Fig. \ref{fig:n+q}.

\section{Q-balls}
\label{sec:qballs}
Q-balls \cite{coleman} are aggregates of squarks $\tilde{q}$, sleptons $\tilde{l}$ and Higgs fields \cite{kuse16}. The scalar condensate inside a Q-ball core has a global baryon number $Q$ (may be also lepton number)
and a specific energy much smaller than 1 GeV/baryon. 
We assume that the $Q$ numbers of quarks and squarks are equal to 1/3 ($Q_q = Q_{\tilde{q}} = 1/3$) or 2/3 ($Q_q = Q_{\tilde{q}} = 2/3$). 
Protons, neutrons and may be electrons can be absorbed in the condensate.
The vacuum expectation value inside a Q-ball core develops along ``flat directions'' of the potential \cite{gerghe}. 
These flat directions are parametrized by combinations of $\tilde{q}$, and $\tilde{l}$ that are electrically neutral. Supposing that the different flavour squarks are not mass degenerate, their numbers inside the Q-ball core would not be equal for the same reason as in the nuclearite core case.
By assuming that the baryon number is packed inside a Q-ball core, one can get upper limits for the Q-ball quantum number $Q$ and for the Q-ball mass $M_Q: Q \le  10^{30}$ and $M_Q \le  10^{25}$ GeV/c$^2$, respectively; Q-balls with $M_Q < 10^8$ GeV/c$^2$ are unstable \cite{kuse18}.

In the early Universe, only neutral Q-balls were produced: Supersymmetric Electrically Neutral Solitons (SENS), which do not have a net electric charge, are generally massive and may catalyse proton decay. 
SENS may obtain an integer positive electric charge absorbing protons in their interactions with matter; thus, we may have Supersymmetric Electrically Charged Solitons (SECS), which have a core electric charge, have generally lower masses and the Coulomb barrier prevents the capture of nuclei. 
SECS have only integer charges because they are colour singlets. Some Q-balls, which have sleptons in the condensate, can also absorb electrons. 

During the propagation in the Universe, SENS may interact with a proton of the interstellar medium, catalyse the proton decay leading to the emission of 2 - 3 pions (or kaons) and transform quarks into squarks via the reaction $qq \rightarrow \tilde{q}\tilde{q}$. Thus, some SENS may become SECS with a charge +1 emitting $2\pi^0$ with total energy emission of about 1 GeV.

When a SENS enters the Earth atmosphere, it could absorb, for example, a nucleus of nitrogen which gives it the positive charge of +7 (SECS with $Z = +7$). The next nucleus absorption is prevented by Coulomb repulsion. If the Q-ball can absorb electrons at the same rate as protons, the positive charge of the absorbed nucleus may be neutralized by the charge of absorbed electrons.
The incoming SENS remain neutral most of the time.
Electrons may be absorbed via the reaction $u + e \rightarrow d + \nu_e$. If, instead, the absorption of electrons is slow or impossible, the Q-ball carries a positive electric charge after the capture of the first nucleus in the atmosphere (SECS). 
Other SENS could ``swallow'' entire atoms (remaining SENS). If a SENS could absorb an electron, it would acquire a negative charge
(SECS with $Z = -1$). 

The Q-balls have been considered as possible cold dark matter candidates; their core sizes should be only one order of magnitude larger than a typical atomic nucleus \cite{kuse18,kuse19}.
The cross section for interaction of SECS with matter depends on the size of the electronic cloud, and SECS should interact with matter in a way not too different from nuclearites, \S \ref{sec:sqm}. 
Thus their energy losses would be about the same as for nuclearites with a constant radius $\sim 10^{-8}$ cm (given by the size of the electronic cloud). 
The experimental limits obtained for nuclearites can be extended to SECS of medium-high masses. 
SECS with $\beta \simeq 10^{-3}$ and $M_Q < 10^{13}$ GeV/c$^2$ could reach an underground detector from above, SENS also from below \cite{kuse18}. 
More details on energy losses of Q-balls in matter are in ref. \cite{MACROnuclea}. Some experimental limits on SECS are shown on the right plot of Fig. \ref{fig:n+q}.

SENS may be detected by their almost continuous emission of charged pions (energy loss of about 100 GeV g$^{-1}$cm$^{2}$). Thus, flux upper limits on SENS could be obtained from limits on $\cm$s, which catalyse proton decay.

\section{Searches in Neutrino Telescopes}
\label{sec:searchesNT}

\subsection{Fast magnetic monopoles}
\label{sec:fast}

\begin{figure*}[tb]
\begin{center}
\includegraphics[width=11.5cm]{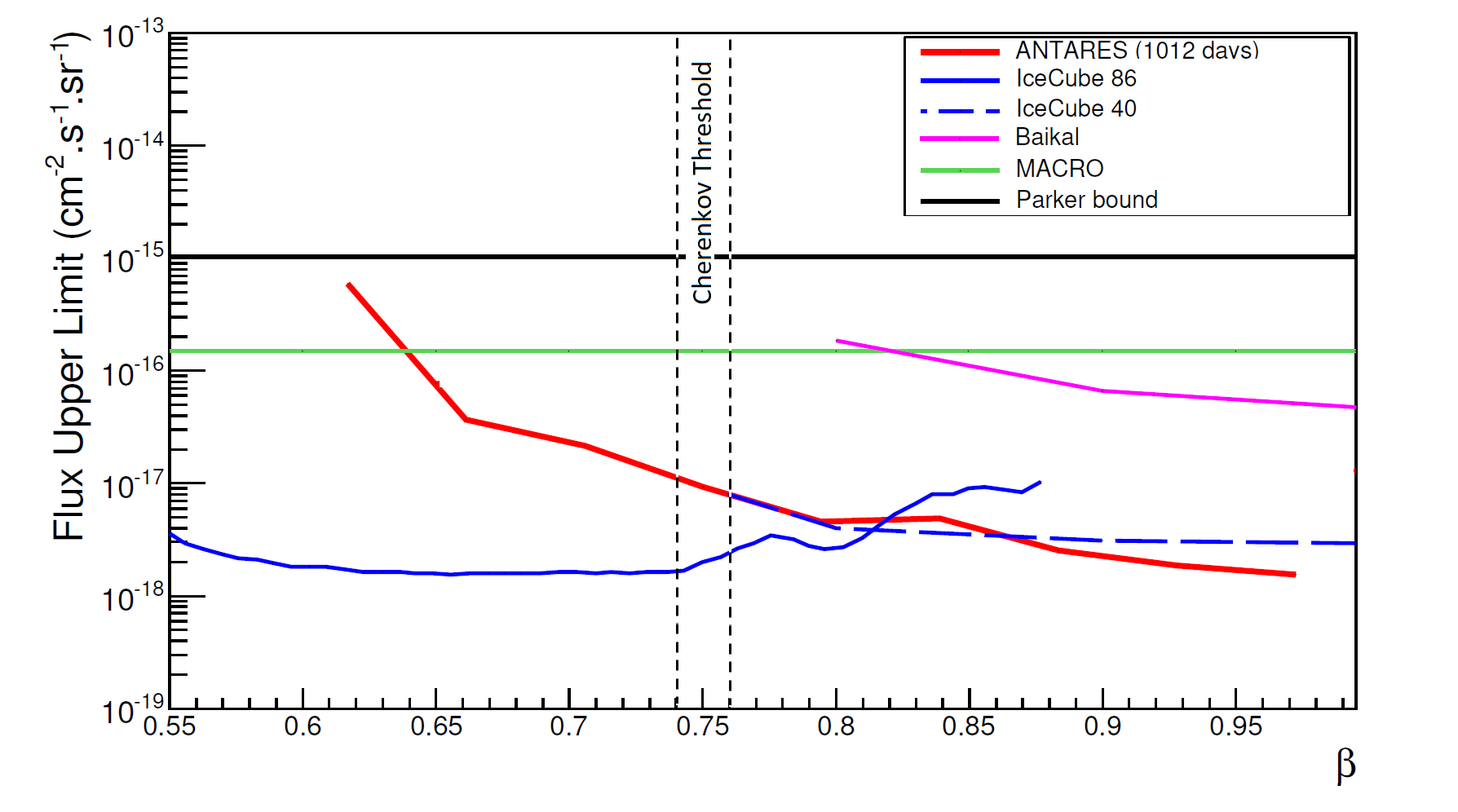}
\end{center}
\caption{\small\label{fig:NTlimits} 90\% C.L. upper limit on flux for $\cm$s in neutrino telescopes in the range $\beta>0.55$. The plot corresponds to a limited region of that presented in Fig. \ref{fig:under}. The MACRO limit (green line) and the Parker bound (black line) are used for benchmark comparison. The vertical lines correspond to the direct Cherenkov threshold in water and ice. Limits for $\beta$ below this threshold rely on the models of $\delta$-ray production. The IceCube limit \cite{ic2} uses the KYG cross section with maximum energy given by Eq. (\ref{eq:tmaxI}). The ANTARES limit uses the Mott cross section and Eq. (\ref{eq:tmaxII}). Thus, the latter is the most conservative, while the IceCube one is the most optimistic, compare with Fig. \ref{fig:drg}. Adapted from \cite{antares2}.}
\end{figure*}
As discussed in \S \ref{sec:cherenkov}, $\cm$s emit direct Cherenkov light in the seawater (ice) when their speed exceeds 0.74c (0.76c). The photons are emitted in a narrow cone at the characteristic Cherenkov angle. Monopoles with such velocities pass through the detector in a straight line and with a constant speed, producing a very large number of photons with respect to that produced by a minimum ionizing particle (compare the cyan line with the black one Fig. \ref{fig:drg}). 
Additional Cherenkov photons are emitted by secondary $\delta$-rays (\S \ref{sec:deltach}); the radiation induced by $\delta$-rays is characterised by a more isotropic emission and can lead to detectable signals of $\cm$s with velocities above $\sim 0.5 c $ (see the discussion on the maximum transferred energy to electrons presented in \S \ref{sec:deltach}). 
Consequently, the time-position correlations between hits in different photomultipliers, and number of detected photoelectrons per hit, caused by a $\cm$ are different from those produced by a muon.

To discriminate signal to background, generally only upgoing events are considered. The solid angle region corresponding to downward going events is dominated by the background of atmospheric muons, with muon bundles \cite{mupage} potentially able to simulate the huge photon yield of relativistic $\cm$s. The Earth acts as a shielding against the muon background.
The possibility that a $\cm$ reaches the detector after passing through the Earth depends on the $\cm$ mass while the Cherenkov emission is independent of the mass and depends on the speed $\beta c$. 
Thus, the detector response of a $\cm$ signal is obtained with simulations focused simply on a benchmark mass without limiting generality. 
This mass value could be $10^{11}$ GeV/c$^2$: all $\cm$s  with mass below this value are expected to be relativistic, see \ref{sec:astroacc}. Monopoles are generated isotropically in the lower hemisphere at different speed above the threshold of $\delta$-ray production. Using these Monte Carlo simulations, the analysis cuts are optimized and sensibilities estimated.

No monopole candidates were identified by the AMANDA \cite{amanda}, Baikal \cite{baikal}, ANTARES \cite{antares1} and IceCube \cite{ic1} experiments.
The latest published results are from update analyses from IceCube \cite{ic2} and ANTARES \cite{antares2}. A summary of the more stringent 90\% C.L. upper limits as a function of $\beta$ is presented in Fig. \ref{fig:NTlimits}. The limits hold in the mass range described by Fig. \ref{fig:KM}: for instance, assuming $\cm$s accelerated by large-scale extragalactic magnetic fields, the limits hold for $10^6 \lesssim M/c^2 \lesssim 10^{14}$ GeV.   

The limits from neutrino telescopes in Fig. \ref{fig:NTlimits} improve the best limit presented in Fig. \ref{fig:under} by about two order of magnitude, but in a limited region of $(\beta , M)$. In the future, the possibility to use the water/ice luminescence (\S \ref{sec:3-lumi}) could extend the limit towards lower values of $\beta$. At present, it is still not completely know the photon yield for Antarctic ice or seawater, and the lifetime of the luminescence processes. In addition, it is not clear if the intrinsic advantage of (almost) null radioactivity background of ice is more important than the light produced by the solutes salts presented in seawater. Finally, the use of luminescence for $\cm$ searches will probably require dedicated trigger logics. 
These studies are extremely important, as neutrino telescopes are the only detectors able to improve significantly the limits well beyond the extended Parker bound (\ref{eq:2.din7}).

\subsection{Monopole-induced proton decay}
\label{sec:searchescata}

In neutrino telescopes, the Cherenkov light from nucleon decays along the $\cm$ trajectory would produce a characteristic hit pattern.
Experimentally, the relevant parameter is the mean free path between two decays
\begin{equation}\label{eq:lcat}
\lambda_{cat}=\frac{1}{\sigma_{cat}\rho}   \ ,
\end{equation}
where $\rho$ is the particle density of the medium through which the $\cm$ propagates. The energy of each cascade, and therefore the number of emitted Cherenkov photons, depends on the decay channel. 
In order to be detected, the light emission due to the catalysis processes must be approximated as being continuous. This condition is satisfied for a mean free path $\lambda_{cat}$ much smaller than the detector spacing. Thus, from an experimental point of view the speed $\beta$ and $\lambda_{cat}$ are the characterizing parameters for these searches.

\begin{figure*}[tb]
\begin{center}
\includegraphics[width=8.5cm]{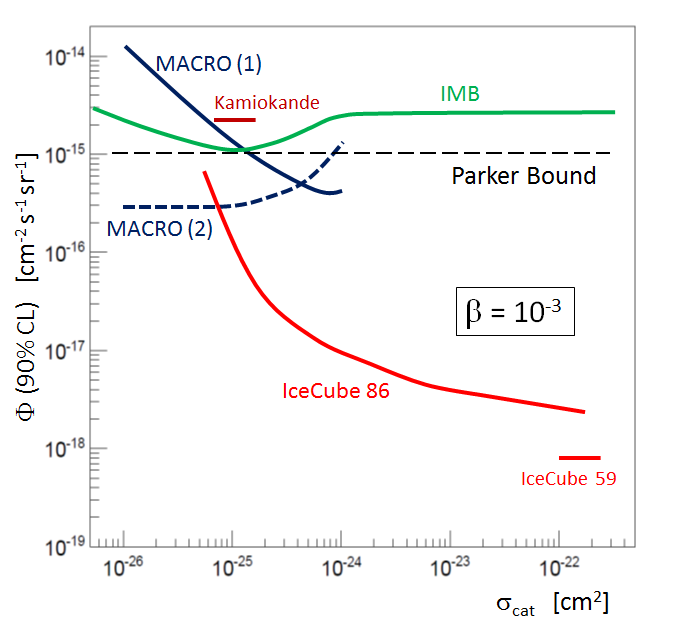}
\end{center}
\caption{\small\label{fig:mcata} Upper limits on the flux of $\beta=10^{-3}$ $\cm$ as a function of the catalysis cross section $\sigma_{cat}$ for two IceCube analyses \cite{ic14}, two MACRO analyses \cite{catalisi}, IMB \cite{MCimb} and Kamiokande \cite{MCkam}.}
\end{figure*}

Early proton decay experiments as Soudan 1 \cite{MCSudan}, IMB \cite{MCimb} and Kamiokande \cite{MCkam} established limits to the $\cm$ flux via the catalysis mechanism.
Additional limits were provided by the underwater detector in Lake Baikal \cite{MCbai,Baikal-cata}.
In the MACRO experiment \cite{catalisi}, a dedicated search based on the streamer tube subsystem was carried out.   
A full Monte Carlo simulation of the expected signal was performed.
The null result of the search was used to set limits on the catalysis cross section, ranging from $10^{-24}$ cm$^2$ to $10^{-26}$ cm$^2$, and correspondingly on the magnetic monopoles flux.
The Super-Kamiokande \cite{SK} collaboration performed an indirect search for GUT monopoles, looking for a neutrino signal coming from proton decay catalyzed by GUT $\cm$s captured in the Sun. 
The IceCube collaboration implemented a dedicated slow-particle trigger in the DeepCore, the central and denser subdetector.
The analysis using one year of data \cite{ic14} yields upper limits (90\% C.L.) for the flux of non-relativistic GUT monopoles at different speeds. 
The limits reported in Fig. \ref{fig:mcata} refer to $\beta=10^{-3}$.

\subsection{Nuclearites and Q-balls}
\label{sec:N+Q}
At present, no neutrino telescope has published results on nuclearite searches, a part preliminary ANTARES sensitivity studies presented at conferences. However, relevant upper limits in a wide nuclearite mass range are possible.

As in other searches for massive particles, the target is for nuclearites gravitationally bound to our Galaxy, with speed $\beta \simeq 10^{-3}$.
For these particles, an upper bound exists given by the mentioned local density of dark matter (see discussion below Eq. \ref{eq:2.phi}); the limit as a function of the nuclearite mass is shown by the red line on the left plot of Fig. \ref{fig:n+q}.
Due to their energy losses (\ref{eq:dedxnuclea}), at that velocity nuclearites with mass $<10^{22}$ GeV/c$^2$ cannot traverse the Earth diameter; thus neutrino telescopes must concentrate their searches to slow, downward going nuclearites.  The characteristic signature against the background of atmospheric muons is their huge energy loss and photon yield. 

\begin{figure*}[tb]
\begin{center}
\includegraphics[width=12.0cm]{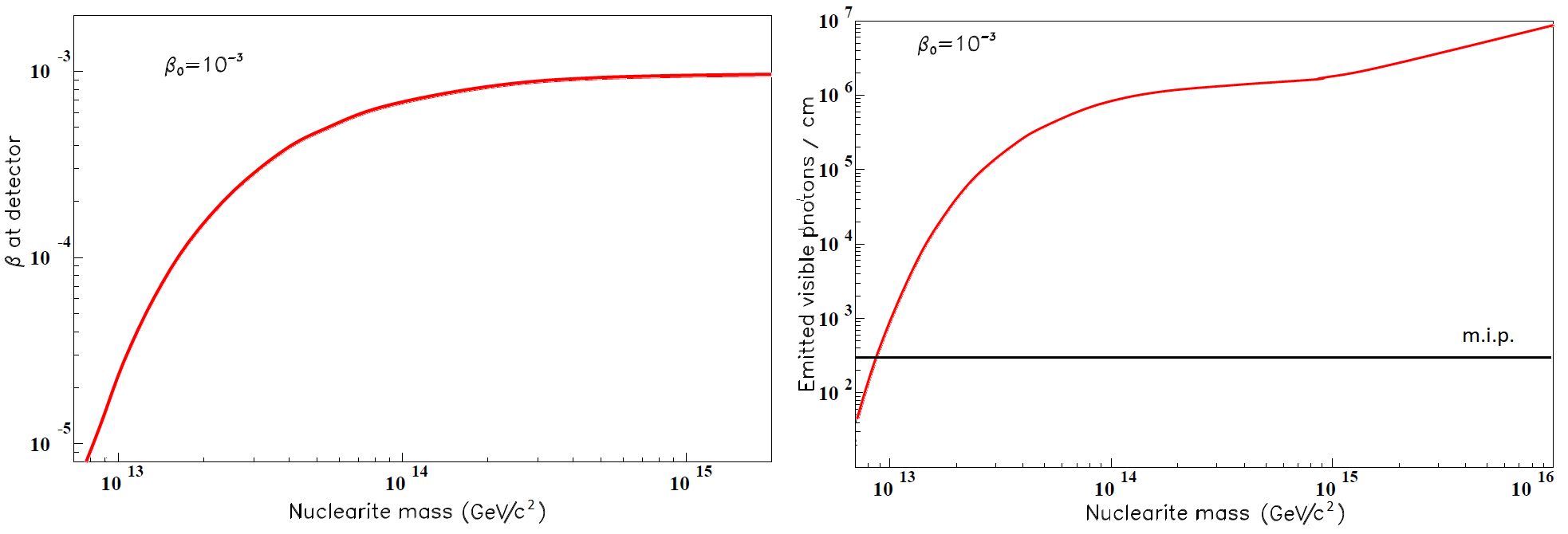}
\end{center}
\caption{\small\label{fig:nucleaNT} 
Left: Nuclearite velocity versus its mass, at depths corresponding to 2100 m.w.e.; the vertically incident nuclearite entered the top of the atmosphere with $\beta = 10^{-3}$.
Right: Number of visible photons emitted per unit path length in water, at 2100 m.w.e. depth, by a vertically downgoing nuclearite. The black line represents the photon yield by a minimum ionizing particle (m.i.p.).}
\end{figure*}

The left panel of Fig. \ref{fig:nucleaNT} shows the velocity versus the nuclearite mass at the levels of the 2100 m of water equivalent, assuming an entry speed in the Earth atmosphere of $\beta = 10^{-3}$ and direction close to the vertical. 
The energy loss is described by Eq. (\ref{eq:dedxnuclea}), assuming the passage in the standard atmosphere (i.e., 10 m.w.e) and 2100 m of medium with density of 1 g/cm$^3$. 
Under the same conditions, the right panel of Fig. \ref{fig:nucleaNT} shows the number of visible photons per cm. As derived in \S \ref{sec:sqm}, a fraction of $3\times 10^{-5}$ of the energy loss is converted into visible light. The change of the slope at mass $\sim 10^{15}$ GeV/c$^2$ is due to the fact that for higher masses the nuclearite size exceeds 1 \AA. 
From those plots, we can derive that the target search is for nuclearites in the mass range above few $10^{15}$ GeV/c$^2$ and below $\sim  10^{20}$ GeV/c$^2$, when the Earth starts to be significantly opaque.

The search strategy could use the fact that a $\beta=10^{-3}$ nuclearite passing close to a photomultiplier (PMT) produce a signal characterized by large widths ($\delta t\sim 30\div 80 \ \mu$s), three orders of magnitude longer than the Cherenkov signals produced by relativistic muons (or intermediate mass monopoles). 
Because the particle travels about 300 m in $ \sim 1$ ms, the detector should observe a succession of similar hits on PMTs displaced along a straight line. These conditions should allow the tracking of the particle along the detector. 
For masses above $10^{17}$ GeV/c$^2$, the photon yield is so large to probably saturate the PMTs working in standard modes. 
Thus, it is mandatory that the data acquisition would:
$i)$ record all hits provided by the nuclearite passage in the detector (some ms, in a km-scale telescope);  
$ii)$ register the time-over-threshold in the PMTs for a signal as long as 100 $\mu$s; and that 
$iii)$ the dynamic range of PMTs allows the acquisition of signals much larger than that produced by minimum ionizing particles. 

Similar techniques are necessary for the searches for other slowly moving downgoing massive particles. For instance, Q-balls can be detected through a process similar to the Rubakov-Callan effect induced by $\cm$s \cite{kasuya}. 
When a Q-ball collides with ordinary matter, nucleons enter the surface layer of the Q-ball, and dissociate into quarks, which are converted to squarks. The Q-ball releases energy of about 1 GeV per collision by emitting soft pions. Therefore, charged particles are created along the path of the Q-ball through a detector.

\section{Conclusions}
\label{sec:conclusions}

The predictions of the Standard Model of particle physics have been spectacularly confirmed by experiments along the last thirty years at accelerators and colliders of increasing energy. 
The final big achievement was the discovery at the CERN LHC of the last missing piece, the Higgs boson.
On the other hand, it is largely believed that the Standard Model is incomplete and represents a sort of low energy limit of a more fundamental and unified theory, which should reveal itself at higher energies.
The energy threshold for this unified interaction is so high that no man-made accelerator, also in the far future, would be able to reach it. 
It is in this context that predictions and searches for magnetic monopoles and other stable massive particles play a fundamental role. 

The search for $\cm$s and that for proton decay, motivated the birth of large underground experiments in the 1980s contributing significantly to the development of astroparticle physics \cite{ms}. 
Although no $\cm$s and no proton decay were observed so far, non-accelerator experiments provided new and unexpected results, exploiting the experimental techniques also used at accelerators in the study of the cosmic radiation, in neutrino physics and astrophysics, in the searches for rare phenomena.
In particular, the lepton sector of the Standard Model was completely revised after the discovery of solar and atmospheric neutrino oscillations by underground detectors. 

The interconnection among cosmic rays studies, neutrino physics and the search for relic massive particles (in particular, $\cm$s) is evident also in future planned experiments, where large area neutrino telescopes and long baseline neutrino experiments can extend the observation in the region of parameters ($M,\beta$) in which stable massive particles can be sought for. 
Examples are the NO$\nu$A detector \cite{nova}, designed to study $\nu_\mu \rightarrow \nu_e$ oscillations in the existing Fermilab NuMI neutrino beam and using liquid scintillator counters, well suited for $\cm$s searches; and the Iron CALorimeter (ICAL) at India-based Neutrino Observatory (INO), for a precise measurement of the neutrino oscillation parameters. Magnetic monopoles will be searched for in ICAL \cite{ino} using resistive plate chambers.
However, these experiments cannot significantly improve the existing upper limits on the $\cm$ and nuclearite flux, as their acceptance is of the order of that of MACRO and SLIM experiments.

Neutrino telescopes cover detection area ($\sim 10^6$ m$^2$) at least three orders of magnitude larger than that of MACRO.  
They can detect relativistic $\cm$s, directly inducing Cherenkov radiation or indirectly from secondary $\delta$-rays, covering the range of speeds $\beta \gtrsim 0.5$. According to astrophysical acceleration models, $\cm$s could be relativistic and cross the Earth in their mass range $10^7 \lesssim Mc^2 \lesssim 10^{14}$ GeV; thus, present limits provided by neutrino telescope refer primarily to IM$\cm$. 
In addition, large area neutrino experiments can catch the light sequence of a $\cm$ catalysing proton decay processes. 
These detectors can reach a sensitivity three order of magnitudes below the Parker bound, but in a very limited interval of $\beta$ or under the assumption of the Rubakov-Callan mechanism.

To possibility to improve the presents limits for GUT monopoles (i.e., in the mass and speed region of $M\sim 10^{17}$ GeV/c$^2$ and $\beta \sim 10^{-3}$) is really challenging. 
If it will be demonstrated that luminescence induced by slowly moving magnetic monopoles in seawater or ice could be exploited, there is the possibility to extend the search to a wider ($M,\beta$) range than the present one.



\begin{thebibliography}{9}

\bibitem{fair07} M. Fairbairn et al.. {"Stable Massive Particles at Colliders."} \emph{Phys. Rept.} {\bf 438}, 1--63, (2007).

\bibitem{ms} M. Spurio. {"Probes of Multimessenger Astrophysics: Charged cosmic rays, neutrinos, $\gamma$-rays and gravitational waves."} Springer (2018). \href{https://doi.org/10.1007/978-3-319-96854-4}{DOI: 10.1007/978-3-319-96854-4}

\bibitem{maxwell} J.C. Maxwell. {"A dynamical theory of the electromagnetic  field."} \emph{Phil. Trans. R. Soc. Lond. } {\bf 155}, 459--512, (1865). 

\bibitem {dirac} P.A.M. Dirac. {"Quantised singularities in the electromagnetic field."} \emph{Proc. Roy. Soc. } {\bf 133}, 60  (1931).

\bibitem{ab} Y. Aharonov, D. Bohm. {"Significance of electromagnetic potentials in quantum theory. "} \emph{Phys. Rev. } {\bf 115}, 485-–491, 
(1959).

\bibitem{Georgi:1972cj} H. Georgi, S.L. Glashow. {"Unified weak and electromagnetic interactions without neutral currents."} \emph{Phys. Rev. Lett. } {\bf 28}, 1494, (1972).

\bibitem{'tHooft:1974qc} G. Hooft't. {"Magnetic Monopoles in Unified Gauge Theories."} \emph{Nucl. Phys. } {\bf B79}, 276, (1974).

\bibitem{Polyakov:1974ek} A.M. Polyakov. {"Particle Spectrum in the Quantum Field Theory."} \emph{JETP Lett. } {\bf 20}, 194, (1974).

\bibitem{Georgi74} H. Georgi, S. Glashow. {"Unity of All Elementary Particle Forces."} \emph{Phys. Rev. Lett. } {\bf 32}, 438, (1974).

\bibitem{pd} K.S. Babu, E. Kearns et al. {"Baryon Number Violation."} \emph{Report of the Community Summer Study (Snowmass 2013), Intensity Frontier -- Baryon Number Violation Group}.  arXiv:1311.5285.

\bibitem{pres84} J. Preskill. {"Magnetic monopoles."} \emph{Ann. Rev. Nucl. Part. Sci.} {\bf 34}, 461, (1984).

\bibitem{Milton} K.A. Milton. {"Theoretical and Experimental Status of Magnetic Monopoles"} \emph{Rept. Prog. Phys.} {\bf 69}, 1637, (2006).

\bibitem{raja} A. Rajantie. {"Introduction to Magnetic Monopoles"} \emph{Contemporary Physics} {\bf 53}, 195, (2012). 

\bibitem{laza-imm} G. Lazarides, C. Panagiotakopoulos, and Q. Shafi. {"Magnetic monopoles from superstring models"} \emph{Phys. Rev. Lett. } {\bf 58}, 1707, (1987).


\bibitem{keph01} T. W. Kephart, Q. Shafi. {"Family Unification, Exotic States and Magnetic Monopoles"} \emph{Phys. Lett. } {\bf B520}, 313--316, (2001).

\bibitem{bhatta} B. Bhattacharjee, G. Sigl. {"Origin and propagation of extremely high-energy cosmic rays."} \emph{Phys. Rept. } {\bf 327}, 109—247, (2000).

\bibitem{pre79} J. Preskill. {"Cosmological Production of Superheavy Magnetic Monopoles."} \emph{Phys. Rev. Lett.} {\bf 43}, 1365,  (1979).

\bibitem{kibble} T.W.B. Kibble. {"Topology of Cosmic Domains and Strings"} \emph{J. Phys.} {\bf A9}, 1387, (1976).


\bibitem{burdin} S. Burdin et al. {"Non-collider searches for stable massive particles."} \emph{Physics Reports} {\bf582}, 1 ,(2015). 

\bibitem{pdg} C. Patrignani et al. (Particle Data Group). {"The Review of Particle Physics."} \emph{Chin. Phys.} {\bf C40}, 100001, (2016).


\bibitem{guth} A. Guth. {"Inflationary universe: A possible solution to the horizon and flatness problems."} \emph{Phys. Rev.} {\bf D23}, 347, (1981).

\bibitem{wick} S. Wick et al. {"Signatures for a Cosmic Flux of Magnetic Monopoles."} \emph{Astropart. Phys. } {\bf 18}, 663, (2003).

\bibitem{parker} E.N. Parker. 
{"The Origin of Magnetic Fields."} \emph{Astrophys. J.} {\bf160}, 383, (1970).

\bibitem{82T1} M.S. Turner, E.N. Parker, and T.J. Bogdan. 
{"Magnetic monopoles and the survival of galactic magnetic fields."} \emph{Phys. Rev.} {\bf D26}, 1296, (1982).


\bibitem{adams93} F. Adams et al. {"Extension of the Parker bound on the flux of magnetic monopoles."} \emph{Phys. Rev. Lett. } {\bf 70}, 2511, (1993).

\bibitem{lewis00} M. J. Lewis, K. Freese, G. Tarl\'e. {"Protogalactic extension of the Parker bound."} \emph{Phys. Rev. } {\bf D62}, 025002, (2000).

\bibitem{ak82} S. P. Ahlen, K. Kinohsita. 
{"Calculation of the stopping power of very-low-velocity magnetic monopoles."} \emph{Phys. Rev. } {\bf D26}, 2347, (1982).


\bibitem{77K1} Y. Kazama, C. N. Yang and A. S. Goldhaber.  {"Scattering of a Dirac particle with charge Ze by a fixed magnetic monopole."} \emph{Phys. Rev. } {\bf D15}, 2287, (1977).


\bibitem{83B1} L. Bracci, G. Fiorentini. {"Binding of magnetic monopoles and atomic nuclei."} \emph{ Phys. Lett. } {\bf B124}, 493, (1983). 

\bibitem{patspu} L. Patrizii, M. Spurio. {"Status of Searches for Magnetic Monopoles."} \emph{ Annu. Rev. Nucl. Part. Sci. } {\bf 65}, 279—302, (2015).

\bibitem {ahlen} S.P. Ahlen. {"Stopping-power formula for magnetic monopoles."} \emph{Phys. Rev. } {\bf D17}, 229, (1978).

\bibitem{derk98} J. Derkaoui et al. {"Energy losses of magnetic monopoles and of dyons in the earth."} \emph{Astropart. Phys. } {\bf 9}, 173, (1998).

\bibitem {geant} S. Agostinelli et al. {"Geant4—a simulation toolkit."} \emph{Nucl. Instr. and Meth.  } {\bf A506}, 250, (2003).

\bibitem{bauer} G. Bauer et al. {"Simulating magnetic monopoles by extending GEANT."} \emph{Nucl. Instr. and Meth. } {\bf A545}, 503, (2005).

\bibitem{bram} B.A.P. van Rens.
{"Detection of magnetic monopoles below the Cherenkov Limit."} PhD thesis, University of Amsterdam, 2006. \url{http://antares.in2p3.fr/Publications}.


\bibitem{tomp} D.R. Tompkins. {"Total Energy Loss and Cherenkov Emission from Monopoles."} \emph{Phys. Rev.} {\bf 138}, B248, (1965). 

\bibitem{Chiarusi} T. Chiarusi and M. Spurio. {"High-Energy Astrophysics with Neutrino Telescopes."} \emph{Eur. Phys. Jour. } {\bf C65}, 649--701, 
(2010).

\bibitem{ah88} S. P. Ahlen. {"Monopole-track characteristics in plastic detectors."} \emph{Phys. Rev. } {\bf D14}, 2935, (1976). 

\bibitem{antares2} A. Albert et al. {"Search for relativistic magnetic monopoles with five years of the ANTARES detector data."} \emph{Journal of High Energy Physics} {\bf 54}, 1707, (2017). 

\bibitem{ic2} M. G. Aartsen et al. {"Searches for Relativistic Magnetic Monopoles in IceCube."} \emph{Eur. Phys. Jour. } {\bf C76}, 133, (2016).

\bibitem{poll} Anna O. Pollmann {"Luminescence of water or ice as a new detection method for magnetic monopoles."} \emph{ Proceedings of the 5th International Conference on New Frontiers in Physics (ICNFP '16)}, arXiv:1610.06397 

\bibitem{ic1} R. Abbasi et al. {"Search for relativistic magnetic monopoles with IceCube."} \emph{ Phys. Rev. } {\bf D87}, 022001, (2013).

\bibitem{81R1}  V.A. Rubakov. {"Adler-Bell-Jackiw anomaly and fermion-number breaking in the presence of a magnetic monopole."} \emph{ Nucl. Phys. } {\bf B203}, 311 (1982).

\bibitem{82C4} C.G. Callan. {"Dyon-fermion dynamics."} \emph{ Phys. Rev.} {\bf D26}, 2058, (1982).


\bibitem{cata_b} J. Arafune, M. Fukugita.  {"Velocity-Dependent Factors for the Rubakov Process for Slowly Moving Magnetic Monopoles in Matter"} \emph{ Phys. Rev. Lett. } {\bf 50}, 1901, (1983). 


\bibitem{nath} P. Nath, P. Fileviez Perez. 
{"Proton stability in grand unified theories, in strings, and in branes."} \emph{ Phys. Rept. } {\bf 441}, 191, (2007).

\bibitem{gps} G. Giacomelli, L. Patrizii and Z. Sahnoun. 
{"Searches for Magnetic Monopoles and ... beyond."} \emph{ Invited paper at the 5th International Conference on "Beyond the Standard Models of Particle Physics, Cosmology and Astrophysics", Cape Town, South Africa, 1 - 6 February 2010}. arXiv:1105.2724 [hep-ex].

\bibitem{cecco} S. Cecchini, M. Spurio. {"Atmospheric muons: experimental aspects."} \emph{Geosci. Instrum. Method. Data Syst.} {\bf 1}, 185, (2012). 

\bibitem{SLIMnuclea} S. Balestra et al. 
{"Results of the search for strange quark matter and Q-balls
with the SLIM experiment."} \emph{ Eur. Phys. Jour.} {\bf C57}, 525, (2008).


\bibitem{gg84} G. Giacomelli. {"Magnetic Monopoles"} \emph{La Rivista del Nuovo Cimento } {\bf v7}, n.12, (1984).

\bibitem{ohya}  S. Orito et al. {"Search for supermassive relics with 2000-m$^2$ array of plastic track detector."} \emph{ Phys. Rev. Lett.} {\bf 66 }, 1951, (1991).	


\bibitem{sou2} J. L. Thron et al. {"A Search for magnetic monopoles with the SOUDAN-2 detector."} \emph{ Phys. Rev. } {\bf D46}, 4846, (1992).

\bibitem{kgf} M. R. Krishnaswamy et al. 
{"Limits on the flux of monopoles from the Kolar gold mine Experiments."} \emph{ Phys. Lett.} {\bf B142 }, 99, (1984).

\bibitem{SLIMmono} S. Balestra et al. {"Magnetic monopole search at high altitude with the SLIM experiment. "} \emph{Eur. Phys. Jour.} {\bf C55 }, 57 (2008).

\bibitem{02A3} M. Ambrosio et al. {"Final results of magnetic monopole searches with the MACRO experiment "} \emph{Eur. Phys. Jour. } {\bf C25}, 511, (2002).

\bibitem{02A4} M. Ambrosio et al. {"The MACRO detector at Gran Sasso."} \emph{ Nucl. Instrum. Meth.} {\bf A486}, 663, (2002).


\bibitem{MAsci} M. Ambrosio et al. {"The Performance of MACRO liquid scintillator in the search for magnetic monopoles with $10^{-3} < \beta < 1$."} \emph{ Astropart. Phys. } {\bf 6}, 113, (1997).  

\bibitem{MAst} M. Ambrosio et al. {"Performance of the MACRO streamer tube system in the search for magnetic monopoles."} \emph {Astropart. Phys. } {\bf 4}, 33, (1995). 

\bibitem{MAall1} M. Ambrosio et al. {"Magnetic monopole search with the MACRO detector at Gran Sasso."} \emph{Phys. Lett. } {\bf B406}, 249, (1997). 

\bibitem{witten} E. Witten. 
{"Cosmic separation of phases."} \emph{Phys. Rev. } {\bf D30}, 272, (1984).

\bibitem{bag} M. Kasuya et al. {"Properties of strange quark matter."} \emph{Phys. Rev. } {\bf D47}, 2153, (1993).

\bibitem{deru}  A. De Rujula, S. L. Glashow. {"Nuclearites: a novel form of cosmic radiation."} \emph{ Nature} {\bf 312}, 734, (1984).

\bibitem{MACROnuclea} M. Ambrosio et al. {"Nuclearite search with the macro detector at Gran Sasso."} \emph{Eur. Phys. Jour.} {\bf C13}, 453, (2000).


\bibitem{coleman} S. Coleman. {"Q-balls."} \emph{ Nucl. Phys.} {\bf B262}, 263--283, (1985).

\bibitem{kuse16} A. Kusenko. 
{"Solitons in the supersymmetric extensions of the Standard Model."} \emph{ Phys. Lett. } {\bf B405}, 108, (1997).

\bibitem{gerghe} T. Gerghetta, C. Kolda, S. P. Martin. 
{"Flat directions in the scalar potential of the supersymmetric standard model."} \emph{Nucl. Phys. } {\bf B468}, 37, (1996).

\bibitem{kuse18} G. Dvali, A. Kusenko, M. Shaposhnikov. {"New physics in a nutshell, or Q-ball as a power plant."}
\emph{Phys. Lett. } {\bf B417}. 99--106, (1998). 

\bibitem{kuse19} A. Kusenko and M. Shaposhnikov. {"Supersymmetric Q-balls as dark matter."}  \emph{ Phys. Lett. } {\bf B418}, 46—54, (1998).


\bibitem{mupage} Y. Becherini, A. Margiotta, M. Sioli and M. Spurio.          
{"A parameterisation of single and multiple muons in the deep water or ice."} \emph{ Astropart. Phys.} {\bf 25}, 1, (2006).  


\bibitem{amanda} R. Abbasi et al. 
{"Search for relativistic magnetic monopoles with the AMANDA-II neutrino telescope."} \emph{ Eur. Phys. Jour. } {\bf C69}, 361, (2010).

\bibitem{baikal} K. Aynutdinov et al. {"Search for relativistic magnetic monopoles with the Baikal Neutrino Telescope."} \emph{ Astropart. Phys. } {\bf 29}, 366, (2008).


\bibitem{antares1} S.~Adri\'an-Mart\'inez et al. 
{"Search for Relativistic Magnetic Monopoles with the ANTARES Neutrino Telescope."} \emph{Astropart. Phys. } {\bf 35}, 634, (2012).  

\bibitem{MCSudan} J. Bartelt et al. 
{"Monopole-flux and proton-decay limits from the Soudan 1 detector."} \emph{Phys. Rev.} {\bf D36}, 1990, (1987).

\bibitem{MCimb} R. Becker-Szendy et al. {"New magnetic monopole flux limits from the IMB proton decay detector."} \emph{Phys. Rev. } {\bf D49}, 2169, (1994).

\bibitem{MCbai} V.A. Balkanov et al. 
{"The Baikal deep underwater neutrino experiment: Results, status, future."} \emph{Prog. Part. Nucl. Phys. } {\bf 40}, 391, (1998). 

\bibitem{MCkam} T. Kajita et al. {"Search for Nucleon Decays Catalyzed by Magnetic Monopoles."} \emph{ J. Phys. Soc. Jpn. } {\bf 54}, 4065, (1985).

\bibitem{Baikal-cata} I. Belolaptikov et al. {"The Baikal underwater neutrino telescope: Design, performance and first results "} \emph{ Astropart. Phys.} {\bf 7}, 263, (1997).

\bibitem{catalisi} M. Ambrosio et al. 
{"Search for nucleon decays induced by GUT magnetic monopoles with the MACRO experiment."} \emph{ Eur. Phys. Jour. } {\bf C26}, 163, (2002).

\bibitem{SK} K. Ueno et al. 
{"Search for GUT monopoles at Super–Kamiokande."} \emph{ Astropart. Phys. } {\bf 36}, 131, (2012). 

\bibitem{ic14} M.G. Aartsen et al. {"Search for non-relativistic Magnetic Monopoles with IceCube."} \emph{ Eur. Phys. Jour. } {\bf C74}, 2938, (2014).


\bibitem{kasuya} S. Kasuya et al. {"IceCube potential for detecting Q-ball dark matter in gauge mediation."} \emph{ Prog. Theor. Exp. Phys. } {\bf 2015}, 053B02, (2015)

\bibitem{nova} Z. Wang et al. {"Software Trigger Algorithms to Search for Magnetic Monopoles with the NOvA Far Detector."} \emph{ J. Phys. Conf. Ser. } {\bf 513}, 012039, (2014).

\bibitem{ino} N. Dash, V. M. Datar, G. Majumder. {"Search for Magnetic Monopole using ICAL at INO."} \emph{ Astropart. Phys.} {\bf 70}, 33--38, (2015). 
\end{thebibliography}
\end{document}